\documentclass{aa}
\usepackage{txfonts}
\usepackage{natbib}
\usepackage{graphicx}
\usepackage{longtable}

\bibpunct{(}{)}{;}{a}{}{,}

\begin{document}

\title{High-resolution X-ray spectroscopy of the low and high states of
  the Seyfert 1 galaxy NGC~4051 with {\it Chandra} LETGS}

\author{K.C.~Steenbrugge\inst{1}
 \and M. Fe\v{n}ov\v{c}\'\i{}k\inst{2}
 \and J.S. Kaastra\inst{2,3}
 \and E.~Costantini\inst{2}
 \and F. Verbunt\inst{3}}

\offprints{K. C. Steenbrugge}

\institute{ 
	 St John's College Research Centre, University of Oxford,
	 Oxford, OX1 3JP, UK 
	 \and
	     SRON Netherlands Institute for Space Research,
Sorbonnelaan 2,
                NL - 3584 CA Utrecht, the Netherlands
         \and
                Astronomical Institute, Utrecht University, P.O. Box
80000,
                NL - 3508 TA Utrecht, the Netherlands
	 }

\date{\today}

\abstract
{ With the new generation of high-resolution X-ray spectrometers the
understanding of warm absorbers in active galactic nuclei has improved
considerably. However, the important questions regarding the distance and
structure of the photoionised wind remain unsolved. }
{ To constrain the distance of the photoionised wind, we
  study the variability of the continuum, absorption, and, emission 
  in one of the brightest and most variable low-luminosity AGN: the
  narrow-line Seyfert galaxy  NGC~4051.}
{ We analyse two observations taken with the Low Energy Transmission
Grating Spectrometer of {\it Chandra}. We investigated the spectral
response to a sudden flux decrease by a factor of 5, which occurred
during the second observation. 
}
{ We detect a highly ionised  absorption component with an outflow velocity of
$-4670$~km\,s$^{-1}$, one of the highest outflow velocity components
  observed in a Seyfert 1 galaxy. Furthermore, this is one of the
    only observations whereby the X-ray observed absorption component
    is unaccompanied by a corresponding UV absorption component with
    the same outflow velocity. The spectra contain a relativistic \ion{O}{viii} Ly$\alpha$ line
with properties similar to those determined for this source with
XMM-{\it Newton}, and four absorption components spanning a range in
ionisation parameter $\xi$ between 0.07 and 3.19 (log values, and
units of 10$^{-9}$ W m).  An emission component producing radiative
recombination continua of \ion{C}{vi} and \ion{C}{v} appears during
the low state. The black body temperature decreases with the drop in
flux observed in the second observation. }
{For all absorber components we exclude that the ionisation parameter
linearly responded to the decrease in flux by a factor of 5. The
variability of the absorber suggest that at least three out of four
detected components are located in the range $0.02-1$~pc. For one
component we only have a lower limit of 0.3~pc. These distances
  are different from earlier suggestions. }


\titlerunning{High-resolution X-ray spectroscopy of NGC~4051}
\authorrunning{K. C. Steenbrugge et al.}

\maketitle

\begin{figure*}[t]
\begin{minipage}{0.5\textwidth}
\includegraphics[width=0.7\textwidth,clip=t,angle=270.]{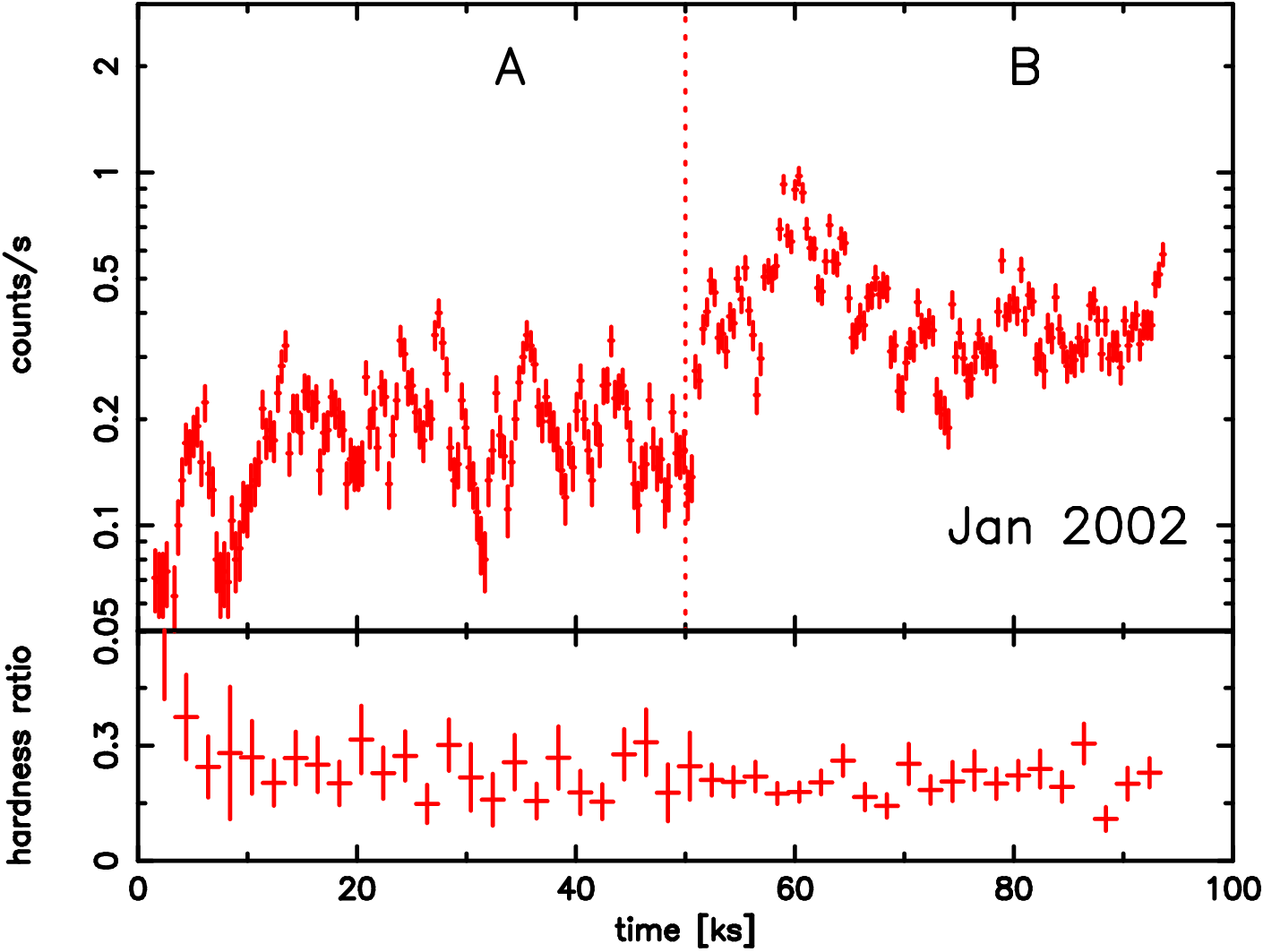}
\end{minipage}
\begin{minipage}{0.5\textwidth}
\includegraphics[width=0.7\textwidth,clip=t,angle=270.]{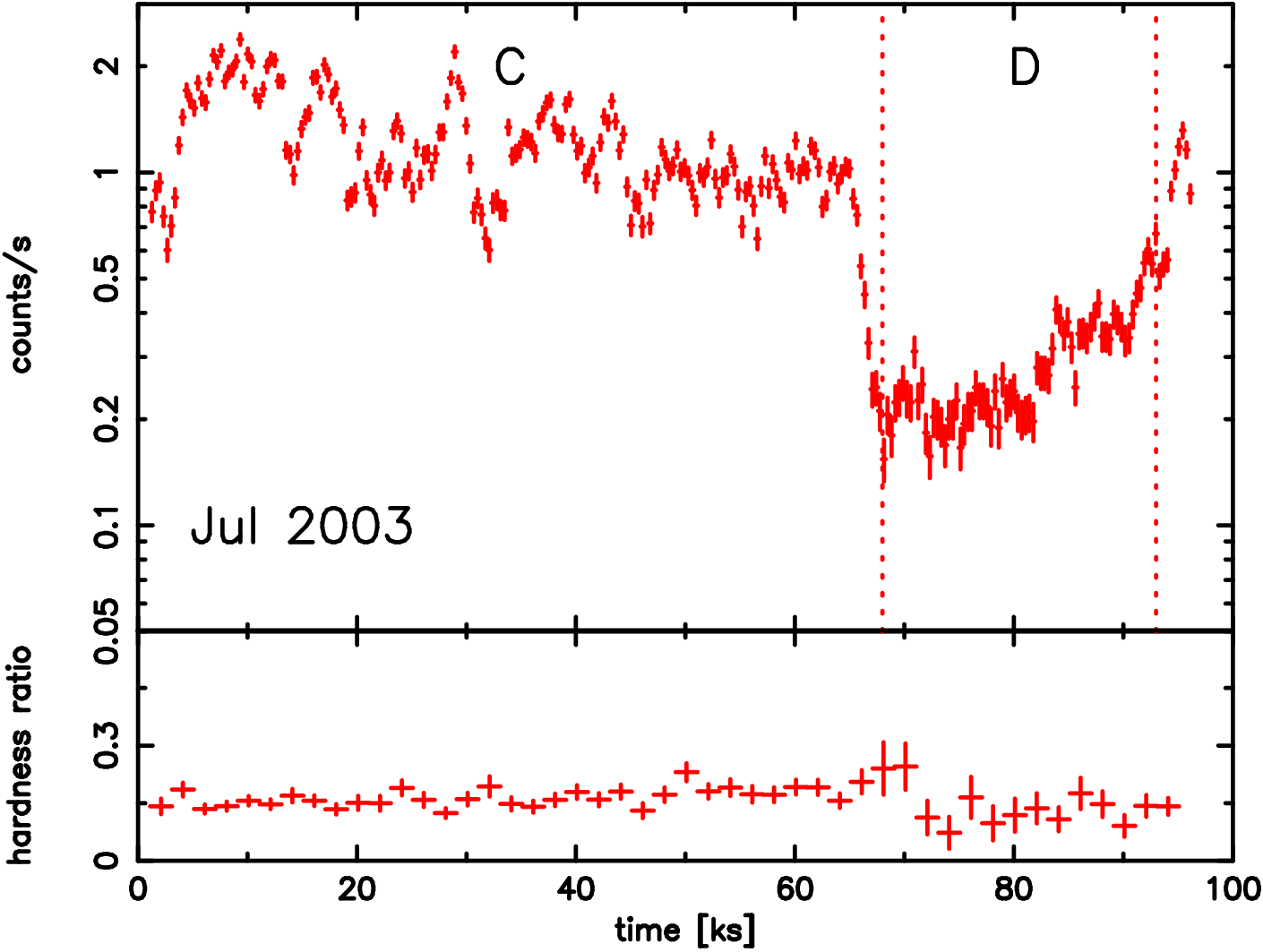}
\end{minipage}
\caption{The light curve in the zeroth spectral order (top panels)
and hardness ratio (bottom panels) for the observations of
Jan 2002 and July 2003. The
hardness ratio is defined as the ratio of the 2 -- 8 and 8 -- 32 \AA{}
counts. The top panel has a bin size of 350~s and the bottom panel has
a bin size of 2000~s. Dotted lines denote the different flux states
and define the intervals used in creating the four spectra.}
\label{fig.lightcurves}
\end{figure*}

\section{Introduction}

The stupendous amount of energy emitted by an active galactic nucleus (AGN) is
released by gas that flows towards a super-massive black hole in the centre of a
galaxy, presumably via an accretion disc. The average  energy of the emitted
radiation increases with decreasing distance to the super-massive black hole, thus X-rays
provide the best probe of the gas flow in the immediate surroundings of the
black hole. From the presence of jets, as well as from blue-shifted ultraviolet
absorption lines, we learn that there is gas flowing away from
the black hole, in addition to the flow towards it. With the availability of X-ray
spectrographs onboard {\it Chandra} and XMM-{\it Newton}, we can study
the lines  emitted in the X-ray band from these outflows close to the
black hole. The presence of multiple absorption line systems that
differ in their level of ionisation or in their outflow velocity
constrains the geometry of the outflow (for example NGC~3783,
\citealt{netzer}; NGC~5548, \citealt{steenbrugge}).    

In this paper we describe the X-ray spectra of the narrow-line Seyfert
1 galaxy NGC\,4051, obtained with the Low-Energy Transmission Grating
(LETG) on board of the {\it Chandra} satellite. Emission from the
nucleus of NGC\,4051 was detected by \citet{hubble}, and the galaxy is
one of the 12 discussed in the seminal paper by \citet{seyfert}. As 
a member of the Ursa Major cluster \citep{tullyetal}, it has a distance
of 18.6~Mpc \citep{tullypierce} and a heliocentric velocity of
700~km\,s$^{-1}$ \citep{verheijen}. NGC\,4051 is a relatively bright
optical object, with $V\simeq13.5$. The soft excess observed with
the Reflection Grating spectrometers (RGS) onboard XMM-{\it Newton} is
not described well by a multi-temperature disc: \citet{ogle}
concludes that the soft excess is due to emission from relativistically broadened
\ion{O}{viii} emission lines and its related radiative recombination
continuum (RRC).  However, \citet{pounds} fitted the EPIC data of the
same observation satisfactorily with a power law plus two black body
components.

Two absorption line systems, at $-2340\pm 130$ and
$-600\pm130$~km\,s$^{-1}$, were detected in the X-ray spectrum with
the High Energy Transmission Grating of {\it Chandra}
\citep{collinge}. Nine absorption systems with a small range in
outflow velocity are detected in the ultraviolet with the Space
Telescope Imaging Spectrograph (STIS) and the Far Ultraviolet
Spectroscopic Explorer (FUSE). One of the absorbers observed in the UV
may coincide with the $-600\pm130\,\mathrm{km\,s}^{-1}$ X-ray
absorption system, all the others have smaller outflow velocities
\citep{collinge,kaspi}.

In a preliminary analysis of our first LETGS spectrum,
\citet{vandermeer} did not detect the $-$2340 km s$^{-1}$ velocity
component, but found indications for the presence of an outflow
component at an even higher velocity of $-4500$~km\,s$^{-1}$. The second
observation, taken when NGC\,4051 was in a much higher flux state,
allows us to investigate the ionised outflow both in terms of outflow
velocity and ionisation structure.  In this paper we report the
analysis of both observations with the LETGS.

\section{Observations and data reduction}

In this paper we describe the two observations of NGC\,4051 taken with
the LETGS \citep{brinkman} onboard of {\it Chandra}. The detector used
was the High-Resolution Camera (HRC-S). The observations span the time
periods December 31, 2001 17:40:41 to January 1, 2002 19:51:46 and July
23, 2003 00:34:15 to July 24 03:24:43, and have a net exposure time of
94.2 and 96.6\,ks, respectively. Henceforth we will we refer to these observations
as the Jan 2002 and Jul 2003 observations. The data reduction method is
described in detail by \citet{kaastra1}. After event
selection we produced background subtracted spectra and light curves.
As the HRC-S detector has limited energy resolution, higher spectral
orders cannot be separated in the data but are taken into account
during spectral fitting. We only use spectra between 2~\AA~and 80~\AA,
as the Galactic absorption attenuates the spectrum strongly above
$\sim 80$~\AA.

\section{Variability}

Fig.~\ref{fig.lightcurves} shows the light curves for each 
observation. For the Jan 2002 observation, the average brightness
of the source increased noticeably in the second half of the exposure. The source was brighter
still during the first $\sim$ 65\,ks of the Jul 2003 observation and then dropped in about 3\,ks to
a lower flux level, which is similar to the flux level of the the
second half of the first observation. The flux recovered
gradually to the high level of the first half of the observation
during the final 20\,ks. On
top of these long-term variations we also detect variability with
smaller, but significant, amplitude on timescales of a few thousand
seconds. These rapid variations have smaller amplitudes after the
rapid flux decrease of the Jul 2003 observation. In the analysis below
we will discuss the spectra at different flux levels, defined by the
lightcurve of Fig.~\ref{fig.lightcurves}: A and B are the first and
second half of the Jan 2002 observation; C and D are the
first and second part of the Jul 2003 observation.

In Fig.~\ref{fig.lightcurves} we also show the hardness ratio, defined
as the ratio between the count rates in the 2-8~\AA\ and 8-32~\AA\
band passes. The hardness variation between different flux levels is
small, with $HR$=0.239 $\pm$ 0.14, 0.203 $\pm$ 0.008, 0.161 $\pm$ 0.003, 0.134
$\pm$ 0.021 for intervals A, B, C and D, respectively with their
1$\sigma$ error. 

\section{Spectral data analysis and results}

\begin{table*}[!htbp]
\begin{center}
\caption{The best fit parameters for the different spectral components
  for the four spectra.  All wavelengths and
  velocities refer to the rest frame of NGC~4051. Numbers in square
  brackets are calculated from the fit parameters; numbers in round brackets
  were kept fixed during the fit.} 
 \begin{tabular}{lcccc}
\hline
\hline
                                         &  A               &   B            &           C    &    D            \\
\hline
\multicolumn{5}{l}{\bf Continuum}\\
\multicolumn{5}{l}{$\bullet\ $\textit{power law}}\\
Flux ($2-10$~keV)$^a$                    &   $0.83\pm 0.07$ & $1.36\pm 0.08$ & $2.93\pm 0.09$ &  $0.48\pm 0.05$ \\
Photon index $\Gamma$                    &   $1.65\pm 0.07$ & $2.00\pm 0.05$ & $2.24\pm 0.02$ &  $2.44\pm 0.05$ \\
\multicolumn{5}{l}{$\bullet\ $\textit{modified black body}}\\   
Flux ($0.2-10$~keV)$^a$                  &   $0.49\pm 0.08$ & $ 0.66\pm0.14$ & $2.0 \pm 0.2 $ &  $0.37\pm 0.09$ \\
$kT$ (eV)                                &   $ 95 \pm  6  $ & $ 96 \pm  8  $ & $143 \pm 4   $ &  $ 90 \pm  5  $ \\
\hline
\multicolumn{5}{l}{\bfseries{Absorption components}}\\
\multicolumn{5}{l}{$\bullet\ $\textit{warm absorber 1}}\\
$N_{\rm H}$ ($10^{25}$~m$^{-2}$)         &        --        &      --        & $0.12\pm 0.04$ &  $  (0.12)    $ \\
$\log\xi$ ($10^{-9}$~W\,m)               &        --        &      --        & $0.07\pm 0.13$ &  $ 0.3\pm 0.2 $ \\
$v_{\rm turb}$ (km\,s$^{-1}$)            &        --        &      --        & $300 \pm  60 $ &  $  (300)     $ \\
$v_{\rm out}$ (km\,s$^{-1}$)             &        --        &      --        & $-210\pm 70  $ &  $  (-210)    $ \\
\multicolumn{5}{l}{$\bullet\ $\textit{warm absorber 2}}\\
$N_{\rm H}$ ($10^{25}$~m$^{-2}$)         &   $  (0.17)    $ & $0.17\pm 0.05$ & $0.29\pm 0.05$ &  $  (0.29)    $ \\
$\log\xi$ ($10^{-9}$~W\,m)               &   $0.95\pm 0.25$ & $0.49\pm 0.20$ & $0.87\pm 0.09$ &  $0.52\pm 0.13$ \\
$v_{\rm turb}$ (km\,s$^{-1}$)            &   $   (230)    $ & $230 \pm 120 $ & $110 \pm  20 $ &  $   (110)    $ \\
$v_{\rm out}$ (km\,s$^{-1}$)             &   $  (-330)    $ & $-330\pm 140 $ & $-200\pm  30 $ &  $    (-200)  $ \\
\multicolumn{5}{l}{$\bullet\ $\textit{warm absorber 3}}\\
$N_{\rm H}$ ($10^{25}$~m$^{-2}$)         &   $   (0.6)    $ & $ 0.6\pm 0.3 $ & $0.8 \pm 0.4 $ &  $  (0.8)     $ \\
$\log\xi$ ($10^{-9}$~W\,m)               &   $2.23\pm 0.17$ & $1.98\pm 0.22$ & $2.32\pm 0.13$ &  $2.4^{+0.6}_{-0.2}$\\
$v_{\rm turb}$ (km\,s$^{-1}$)            &   $   (100)    $ & $ 100\pm 50  $ & $ 90 \pm  30 $ &  $   (90)     $ \\
$v_{\rm out}$ (km\,s$^{-1}$)             &   $ (-610)     $ & $-610\pm 110 $ & $-580\pm 50  $ &  $   (-580)   $ \\
\multicolumn{5}{l}{$\bullet\ $\textit{warm absorber 4}}\\
$N_{\rm H}$ ($10^{25}$~m$^{-2}$)         &   $    (9)     $ & $  9 \pm 5   $ & $ 20 \pm 10  $ &  $   (20)     $ \\
$\log\xi$ ($10^{-9}$~W\,m)               &   $3.06\pm 0.10$ & $2.92\pm 0.17$ & $3.19\pm 0.09$ &  $ 3.2\pm 0.2 $ \\
$v_{\rm turb}$ (km\,s$^{-1}$)            &   $   (120)    $ & $120 \pm 80  $ & $ 19 \pm  13 $ &  $   (19)     $ \\
$v_{\rm out}$ (km\,s$^{-1}$)             &   $ (-4590)    $ & $-4590\pm 180$ & $-4670\pm 150$ &  $  (-4670)   $ \\
\hline
\multicolumn{5}{l}{\bfseries{Emission lines}}\\
\multicolumn{5}{l}{$\bullet\ $\textit{\ion{O}{viii} Ly$\alpha$ (relativistic)$^b,c$}} \\
Flux$^d$                                 &   $ 12 \pm 2   $ & $ 22 \pm 5   $ & $ 6.8\pm 1.5 $  & $ 3.5\pm 1.1 $ \\
$q$ (index Laor profile)                 &   $    (5.2)   $ & $ 5.2\pm 0.5 $ & $2.06\pm 0.26$  & $  (2.06)    $ \\
FWHM (\AA)                               &   $    [10]    $ & $   [10]
$ & $  [3.0]     $  & $  [3.0]     $ \\\hline
\multicolumn{5}{l}{$\bullet\ $\textit{\ion{O}{viii} Ly$\alpha$ (BLR)$^b$}} \\
Flux$^d$                                 &   $ 0.1\pm 0.2 $ & $ 0.0\pm 0.2 $ & $ 1.8\pm 0.6 $  & $ 0.8\pm 0.4 $ \\
FWHM (\AA)                               &   $   (0.45)   $ & $  (0.45)    $ & $0.45\pm 0.14$  & $  (0.45)    $ \\
\multicolumn{5}{l}{$\bullet\ $\textit{\ion{O}{vii} triplet (BLR)}} \\
Flux$^d$                                 &   $ 1.4\pm 0.6 $ & $ 0.0\pm 0.7 $ & $ 6.2\pm 1.4 $  & $ 2.2\pm 1.2 $ \\
FWHM (\AA)                               &   $   (1.7)    $ & $  (1.7)     $ & $ 1.7\pm 0.4 $  & $  (1.7)     $ \\
$\lambda$ (\AA)                          &   $  (22.11)   $ & $ (22.11)    $ & $22.11\pm 0.13$ & $ (22.11)    $ \\
\multicolumn{5}{l}{$\bullet\ $\textit{\ion{O}{vii} forbidden (narrow)}} \\
Flux$^d$                                 &   $0.8 \pm 0.2 $ & $ 0.9\pm 0.2 $ & $ 1.0\pm 0.2 $  & $ 1.3\pm 0.3 $ \\
$\lambda$ (\AA)                          &$22.087\pm 0.005$&$22.081\pm 0.006$&$22.090\pm 0.009$& $  (22.090)  $ \\
\multicolumn{5}{l}{$\bullet\ $\textit{\ion{C}{vi} Ly$\alpha$ (BLR)$^e$}} \\
Flux$^d$                                 &   $ 0.6\pm 0.4 $ & $ 1.0\pm 0.5 $ & $ 2.4\pm 0.8 $  & $ 3.1\pm 0.7 $ \\
FWHM (\AA)                               &   $   (0.36)   $ & $ (0.36)     $ & $0.36\pm 0.13$  & $  (0.36)    $ \\
\hline
\end{tabular}
\\
\label{tab:fit}
\noindent
$^{a}$ Absorption-corrected flux in 10$^{-14}$ W\,m$^{-2}$. \\
$^{b}$ Wavelength is frozen to the rest wavelength of 18.967 \AA~(for
\ion{O}{viii} Ly$\alpha$). \\ 
$^{c}$ And an inclination angle of 48$^{\circ}$. \\
$^{d}$ Absorption-corrected photon flux in photons\,m$^{-2}$\,s$^{-1}$. \\
$^{e}$ Wavelength is frozen to the rest wavelength of 33.736 \AA. \\
\end{center}
\end{table*}

From the above (Fig.~\ref{fig.lightcurves}) it is clear that NGC~4051 shows significant luminosity
variability during our observations. As the level of the variations is
quite strong (e.g., a factor of five in flux between parts C and D of
the July 2003 observation), it is not useful to analyse the integrated
spectra as a whole, but we will rather analyse spectra A, B, C and D
separately, with a focus on the spectral differences between these
parts. Statistics prohibit us to do spectral analysis down to the
shortest variability timescales of less than 1000~s.

For all spectral fitting we used the SPEX code developed and
  regularly updated with new models and atomic data at SRON
\citep{kmn}\footnote{http://www.sron.nl/spex}. As the flux is low in some of the spectra, we cannot use
$\chi^2$ minimisation, instead we minimise the C-statistic. Note
that that the uncertainty in absolute calibration for the {\it
Chandra} LETGS is 13~m\AA. Therefore a velocity difference smaller
than $\sim$ 200~km~s$^{-1}$ between spectra A/B and C/D is unlikely
due to an intrinsic velocity shift.

For our spectral model we included different components: a broad-band
continuum, narrow and broad emission features, and photoionised
absorption. The need for these components and the resulting
parameters will be discussed separately.

Noting that spectrum C has the highest signal-to-noise of the four spectra, we
decided to fit this spectrum first and then use the best fit model as
a template for fitting spectrum B. The signal-to-noise ratios for spectra A
and D are poor, therefore we only left a limited number of parameters
free when fitting these spectra.

\begin{table}[!htbp]
\begin{center}
\caption{Significances of the different spectral components. We list
  the increase  in the C-statistic when the component is omitted from
  the model and the spectrum is refitted.}  
 \begin{tabular}{lcc}
\hline
\hline
Component                     & Spectrum B & Spectrum C \\
Modified black body           & 7.36       & 88.97 \\
rel. \ion{O}{viii}            & 16.84      & 11.32 \\
BLR \ion{O}{viii}             & 0          & 16.69 \\
BLR \ion{O}{vii}              & 0          & 31.49 \\
narrow \ion{O}{vii} forbidden & 36.08      & 24.60 \\
BLR \ion{C}{vi}               & 4.07       & 12.01 \\
warm absorber 1               & --         & 14.53 \\
warm absorber 2               & 12.74      & 17.83 \\
warm absorber 3               & 12.75      & 14.55 \\
warm absorber 4               & 45.19      & 32.89 \\
\hline
\label{tab:delta_c}
\end{tabular}
\end{center}
\end{table}

The best fit parameters for our spectral model are listed in Table~\ref{tab:fit}. All models
were corrected for Galactic absorption with a column density of
$1.31\times 10^{24}$~m$^{-2}$ \citep{elvis} and for the cosmological
redshift (including peculiar motion) of 700~km\,s$^{-1}$. All errors
correspond to $\Delta C = 1$ (68~\% confidence). The best fit values
for the C-statistics are 2700 (2476), 2658 (2466), 2738 (2599), and
2870 (2610) for components A--D; the numbers in brackets are the
degrees of freedom. In Table~\ref{tab:delta_c} we show the increase
$\Delta C$ of the C-statistic when a given spectral component is
omitted from the model (and re-fitting the spectrum; the omission of a
given component can sometimes be partly compensated by adjusting the
parameters of the other components). We did this for spectra B and C,
the best quality spectra. 

\begin{table}[!htbp]
\begin{center}
\caption{The best fit parameters for the RRCs in spectrum D.  The
  velocity refers to the rest frame of NGC~4051.}  
 \begin{tabular}{lc}
\hline
\hline
Parameter & value \\
$kT$ (eV)              & $    5\pm  2  $ \\
\ion{C}{v}$^{a}$       & $  210\pm 110 $ \\
\ion{C}{vi}$^{a}$      & $  390\pm  80 $ \\
\ion{N}{vi}$^{a}$      & $   40\pm  60 $ \\
\ion{N}{vii}$^{a}$     & $   30\pm  30 $ \\
\ion{O}{vii}$^{a}$     & $    0\pm  20 $ \\
\ion{O}{viii}$^{a}$    & $   10\pm  14 $ \\
velocity (km\,s$^{-1}$) & $+1300\pm 200 $ \\
\hline
\end{tabular}
\\
\label{tab:fitrrc}
\noindent
$^{a}$ Emission measures in $10^{64}$~m$^{-3}$, e.g., X=\ion{C}{vii} produces
the RRC observed for \ion{C}{vi}.\\
\end{center}
\end{table}

\begin{figure*}[!htbp]
\begin{center}
\includegraphics[angle=-90,width=\hsize,totalheight=11cm]{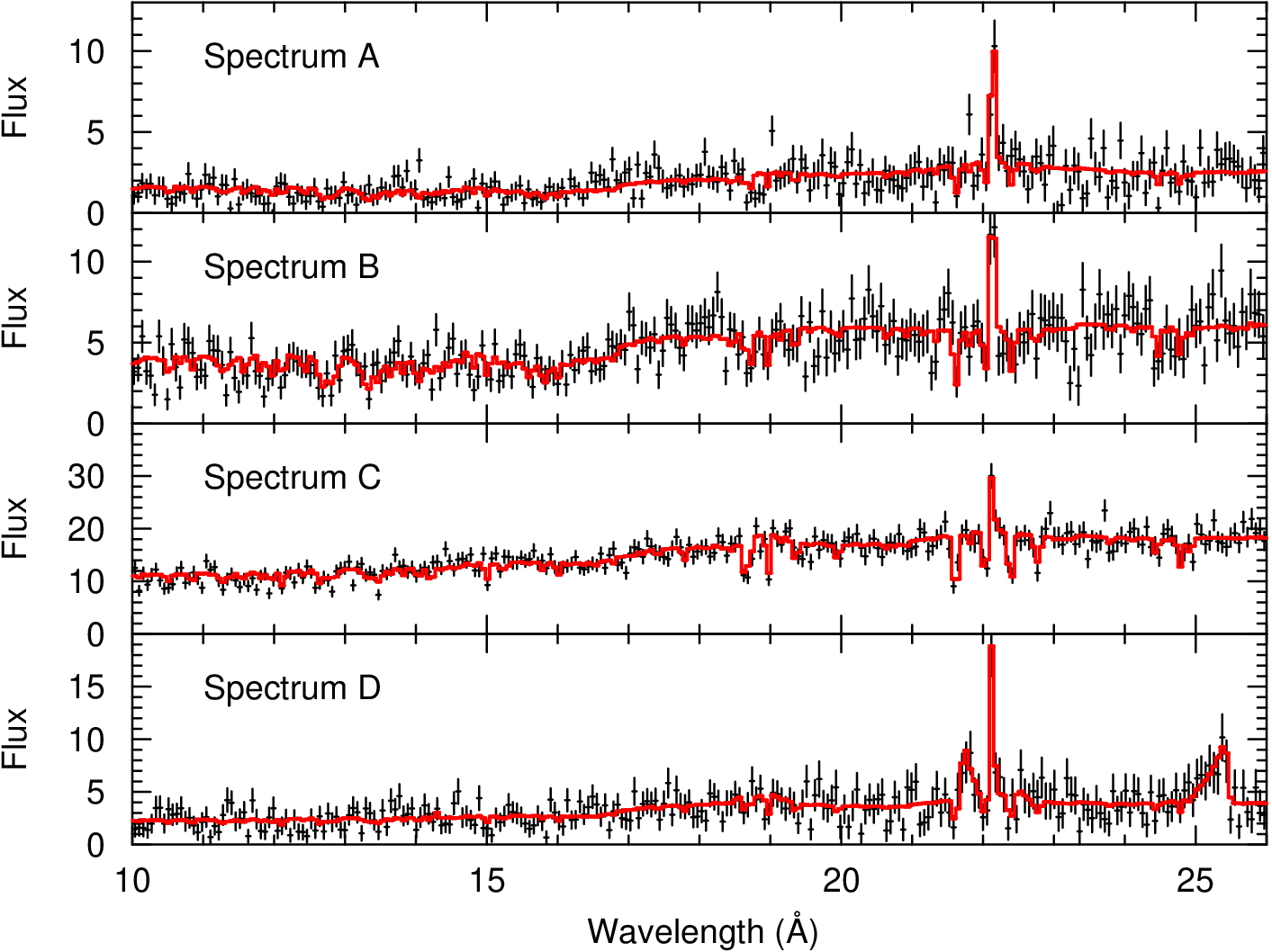}
\vskip0.5cm
\includegraphics[angle=-90,width=\hsize,totalheight=11cm]{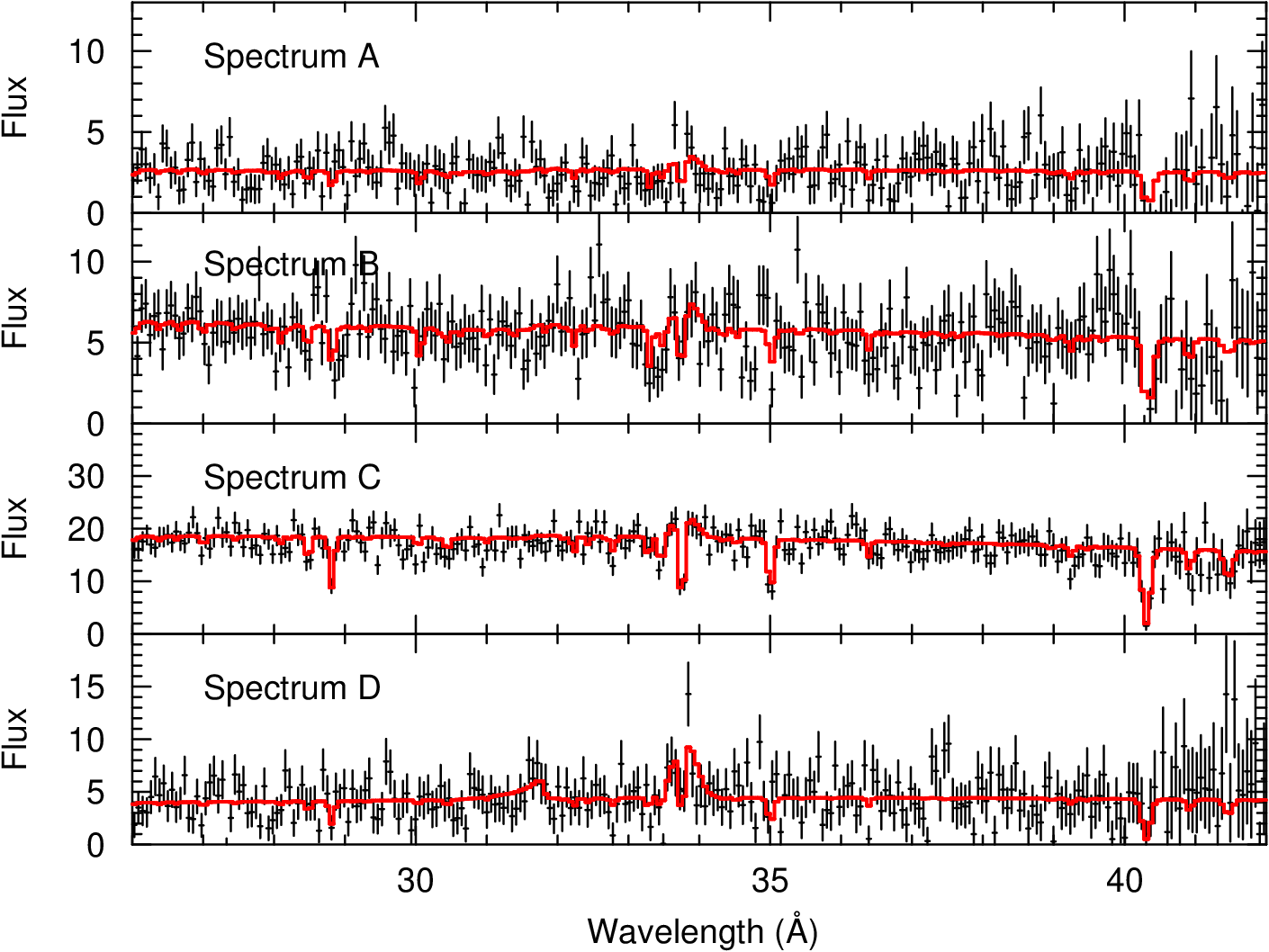}
\caption{Flux (in units of photons m$^{-2}$ s$^{-1}$ \AA$^{-1}$)
  and our best spectral model.  All spectra are binned by a
  factor of 2 for clarity of display. Note the different continuum
  levels and shapes between the spectra, especially around the upturn
  at $\sim$ 16~\AA. In spectrum D 
  \ion{C}{vi} and weaker \ion{C}{v} RRCs at 25.43 \AA\ and 31.80
  \AA\ respectively, are observed. }
\label{fig:spectrum_ABCD}
\end{center}
\end{figure*}

Finally, in spectrum D we afterwards added a spectral component
fitting the radiative recombination continua (RRCs). Only the
\ion{C}{v} and \ion{C}{vi} RRCs are present, and can be identified by
eye in Fig.~\ref{fig:spectrum_ABCD}. The temperature and emission
measures $n_{\rm e}n_{\rm X}V$ with $V$ the emitting volume, $n_{\rm
e}$ the electron density, $n_{\rm X}$ the density of the recombining
ion X giving rise to the observed transitions are given in
Table~\ref{tab:fitrrc}.

\subsection{Continuum}

In spectrum C the continuum cannot be fitted with only a power
law. The soft excess can be modelled by adding a black body modified
by coherent Compton scattering \citep{kaastra2}. Examining all
spectra, we find that the power-law photon index becomes softer as
time progresses, and is thus not a simple function of luminosity. The
modified black body temperature seems to track the luminosity
variations; the temperature is highest for the high state and
drops back to a lower value after the drop in flux just before the
start of spectrum D.

\subsection{The ionised absorber}

Thanks to the excellent spectral resolution of the LETG instrument we
can study the warm absorber in detail. We fitted the warm absorber
with the {\it xabs} model in SPEX. In this model, the ionic column
densities are not independent quantities, but are linked via the
ionisation parameter $\xi=L/nr^2$, where $L$ is the source luminosity,
$n$ the hydrogen density and $r$ is the distance from the ionising
source. For the spectral energy distribution of the ionising source we used
the standard one included in SPEX (see dot-dash line of figure 1
    in \citealt{steenbrugge}). In addition to the ionisation
parameter, the fit parameters are the hydrogen column density of the
absorber $N_H$, the outflow velocity $v_{out}$, and the turbulent
velocity $v_{turb}$. The advantage of the {\it xabs} model is that all
relevant ions are taken into account including also the ones with the
weakest absorption features. In the fit the abundances were left
frozen to the solar abundances given by \cite{anders89}.

\begin{figure}[!htb]
\begin{center}
\includegraphics[angle=-90,width=0.45\textwidth]{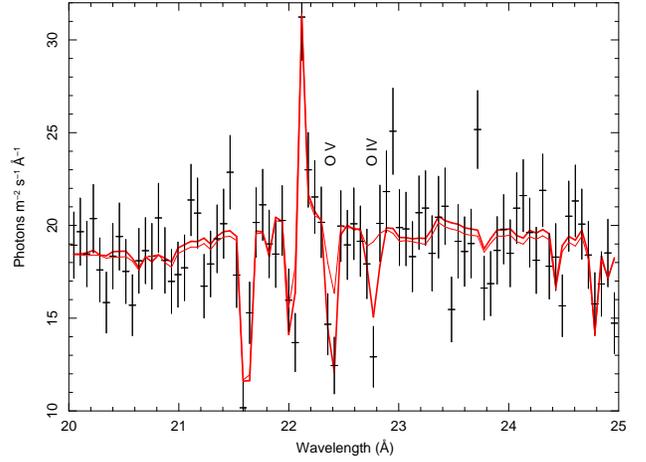}
\caption{Detail of spectrum C fitted with 3 warm absorber components
  (thin line) and 4 warm absorber components (thick line). Binning as
  in Fig.~\ref{fig:spectrum_ABCD}. The best fit 3 component model does
  under-predict the depth of the \ion{O}{v} and \ion{O}{iv} lines.}
\label{fig:4comp}
\end{center}
\end{figure}

Table~\ref{tab:fit} lists the best fit values for the four different
spectra. To adequately describe the warm absorber in spectrum C we
need 4 components. The ionisation parameters range between $0.07$ and
$3.19$ (logarithmic values in 10$^{-9}$~W~m, used throughout this
paper).  Adding a 5$^{th}$ absorption component for spectrum C
does not significantly improve the fit, namely $\Delta$~C decreases by
only 1.91. Fig.~\ref{fig:4comp} compares the best fit model with 3
and 4 absorption components and Table~\ref{tab:fit3comp} gives
the best fit parameters for the three-absorber-component fit to spectrum
C. The fit with 4 warm absorber components better fits the \ion{O}{v}
and \ion{O}{iv} absorption lines. In the remaining three spectra due
to the lower signal to noise ratios, not all four absorption
components are detected.

\begin{table}[!htbp]
\begin{center}
\caption{The best fit parameters for a three-absorber-component model
  for spectrum C. Components $1 - 3$ correspond to components $2 - 4$
  in Tables~\ref{tab:fit} and \ref{tab:delta_c}.}  
 \begin{tabular}{lc}
\hline
\hline
Parameter & value \\
\multicolumn{2}{l}{$\bullet\ $\textit{power law}}\\
Flux ($2-10$~keV)$^a$   & 2.86 $\pm$ 0.04\\
Photon index $\Gamma$ & 2.21 $\pm$ 0.02 \\
\multicolumn{2}{l}{$\bullet\ $\textit{modified black body}}\\ 
flux ($0.2-10$~keV)$^a$ & 1.31 $\pm$ 0.14 \\
$kT$ (eV) & 140 $\pm$ 4 \\
\hline
\multicolumn{2}{l}{\bfseries{Absorption components}}\\
\multicolumn{2}{l}{$\bullet\ $\textit{warm absorber 1}}\\
N$_{\rm H}$ ($10^{25}$~m$^{-2}$) & 0.35 $\pm$ 0.14 \\
$\log\xi$ ($10^{-9}$~W\,m)  & 0.73 $\pm$ 0.06 \\
$v_{\rm turb}$ (km\,s$^{-1}$)    & 147 $\pm$ 21 \\
$v_{\rm out}$ (km\,s$^{-1}$)   & -213 $\pm$ 27\\
\multicolumn{2}{l}{$\bullet\ $\textit{warm absorber 2}}\\
N$_{\rm H}$ ($10^{25}$~m$^{-2}$) & 0.72 $\pm$ 0.3 \\
$\log\xi$ ($10^{-9}$~W\,m) & 2.27 $\pm$ 0.06\\
$v_{\rm turb}$ (km\,s$^{-1}$)   & 100 $\pm$ 28 \\
$v_{\rm out}$ (km\,s$^{-1}$)  & -548 $\pm$ 45 \\
\multicolumn{2}{l}{$\bullet\ $\textit{warm absorber 3}}\\
N$_{\rm H}$  ($10^{25}$~m$^{-2}$) & 23 $\pm$ 7.5 \\
$\log\xi$ ($10^{-9}$~W\,m) & 3.13 $\pm$ 0.04 \\
$v_{\rm turb}$ (km\,s$^{-1}$) & 8 $\pm$ 8 \\
$v_{\rm out}$ (km\,s$^{-1}$) & -4530 $\pm$ 80 \\
\hline
\end{tabular}
\\
\label{tab:fit3comp}
\noindent
$^{a}$ Absorption-corrected flux in 10$^{-14}$ W\,m$^{-2}$. \\
\end{center}
\end{table}

For spectrum B, we need three components. By comparing the parameters of
these components to the parameters of the components for spectrum C, we find
that component 1, which has the smallest hydrogen column density,
is lacking in spectrum B. 

The statistical quality of spectra A and D are much poorer; in fitting
these spectra, we fixed several parameters to the best-fit
values obtained for spectra B and C, respectively. It is quite
reasonable to keep the outflow and turbulent velocity constant between
observations A/B and also between C/D. When we leave both the column
density and the ionisation parameter free in spectrum A, only
component 2 is marginally detected. These non-detections do not imply that the warm absorber
has disappeared, but merely that the statistical quality of 
spectra A and D are insufficient to determine its parameters with
accuracy. For this reason, we make the additional assumption that the
column density of the absorber does not change. Any changes
  however, are likely to be in the ionization parameter (that is, if it can
respond fast enough to the continuum flux variations).

Comparing spectra C and D, only for component
2 a $2.3\sigma$ drop in ionisation parameter is found; the other
components are consistent with no change in the warm
absorber. Comparing spectra A and B, we find only marginally higher
ionisation parameters for components $2-4$ in spectrum A.

\begin{figure}[!htbp]
\begin{center}
\resizebox{\hsize}{!}{\includegraphics[angle=-90]{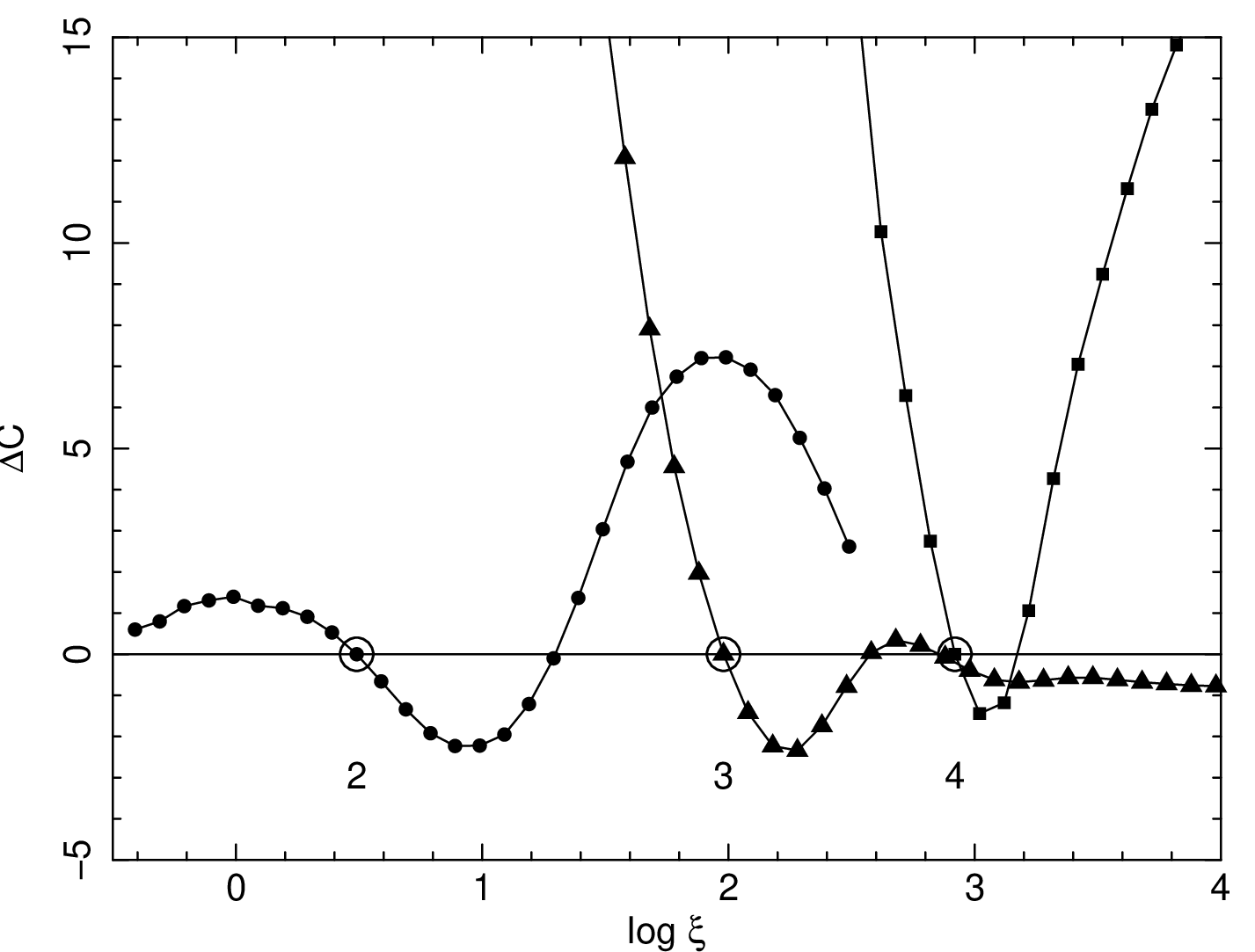}}
\resizebox{\hsize}{!}{\includegraphics[angle=-90]{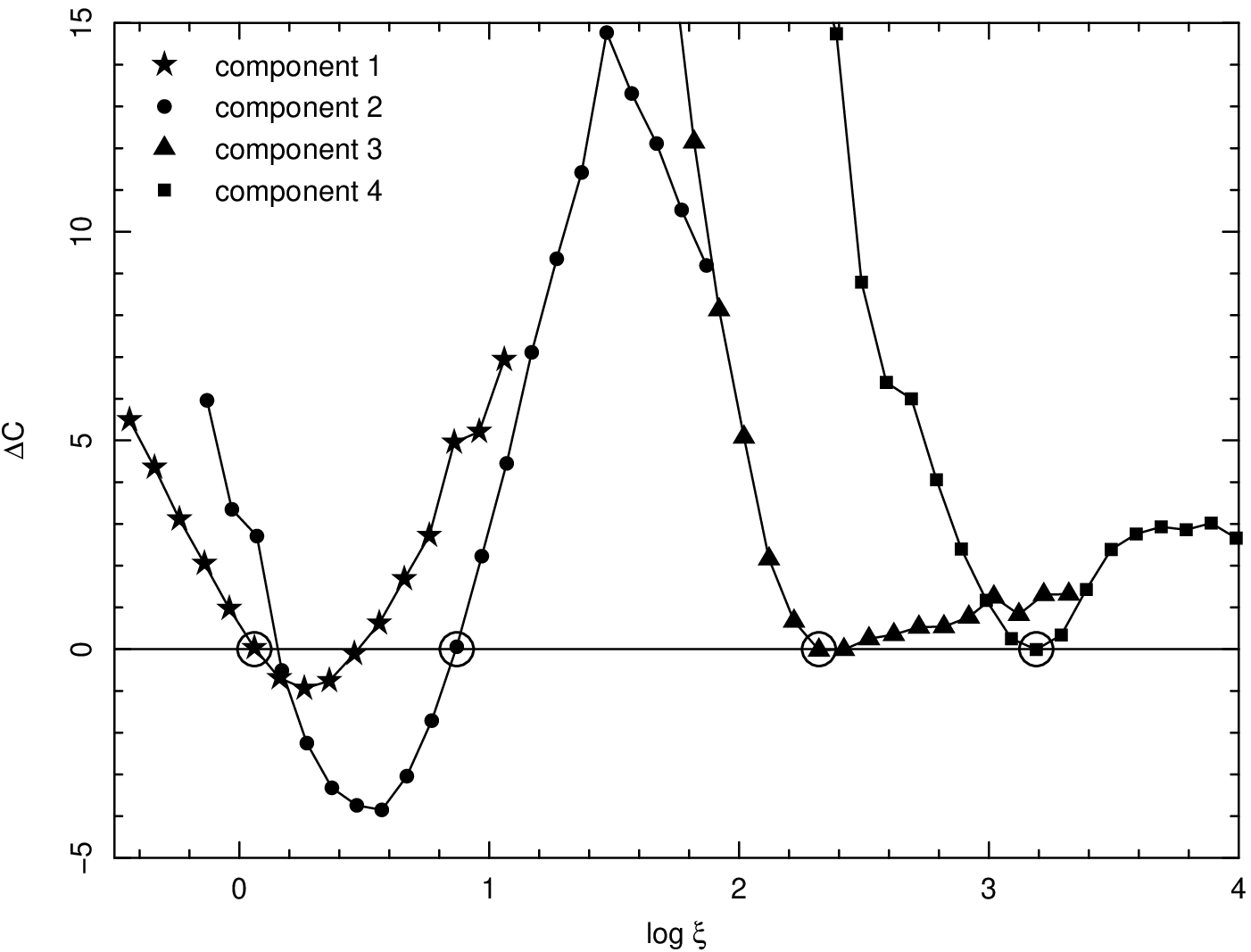}}
\caption{Change ($\Delta C$) in C-statistic for spectrum A (top panel)
and spectrum D (bottom panel) when the ionisation parameter $\xi$ 
of each component is varied. Open circles indicate the reference values of 
each component (for spectrum D, the value of $\xi$ is the one determined from 
spectrum C; for spectrum A, the value of $\xi$ is the one determined from spectrum B).}
\label{fig:stepad}
\end{center}
\end{figure}

Fig.~\ref{fig:stepad} shows the change in C-statistics for spectra A
and D when the ionisation parameter for each component is separately
varied. Step size is 0.1 in log $\xi$ and at each step the goodness
of fit is calculated. It is clear from this figure that no significant changes can
be detected for components 1, 3 and 4.  We discuss the variability of
component 2 in more detail later.

\begin{figure}[!htb]
\begin{center}
\includegraphics[angle=-90,width=0.45\textwidth]{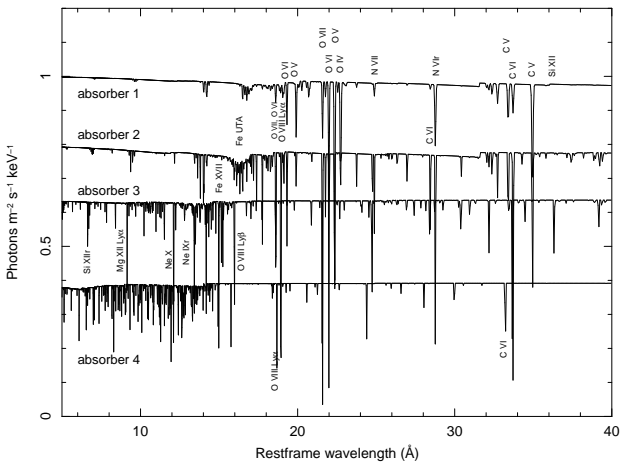}
\caption{The model absorption spectra for each of the 4 different
  absorption components as detected in spectrum C. The continuum was
  modelled to be flat and was renormalised
  per component.}  
\label{fig:comp1}
\end{center}
\end{figure}

In Fig.~\ref{fig:comp1} we present the model absorption spectrum
  for
each component, labelling the deepest lines. It thus allows the reader
to see which component contributes to which line. The continuum was
renormalised to ease identification.

\subsection{Broad emission lines}

\begin{figure*}[!htb]
\begin{center}
\includegraphics[angle=-90,width=0.90\textwidth]{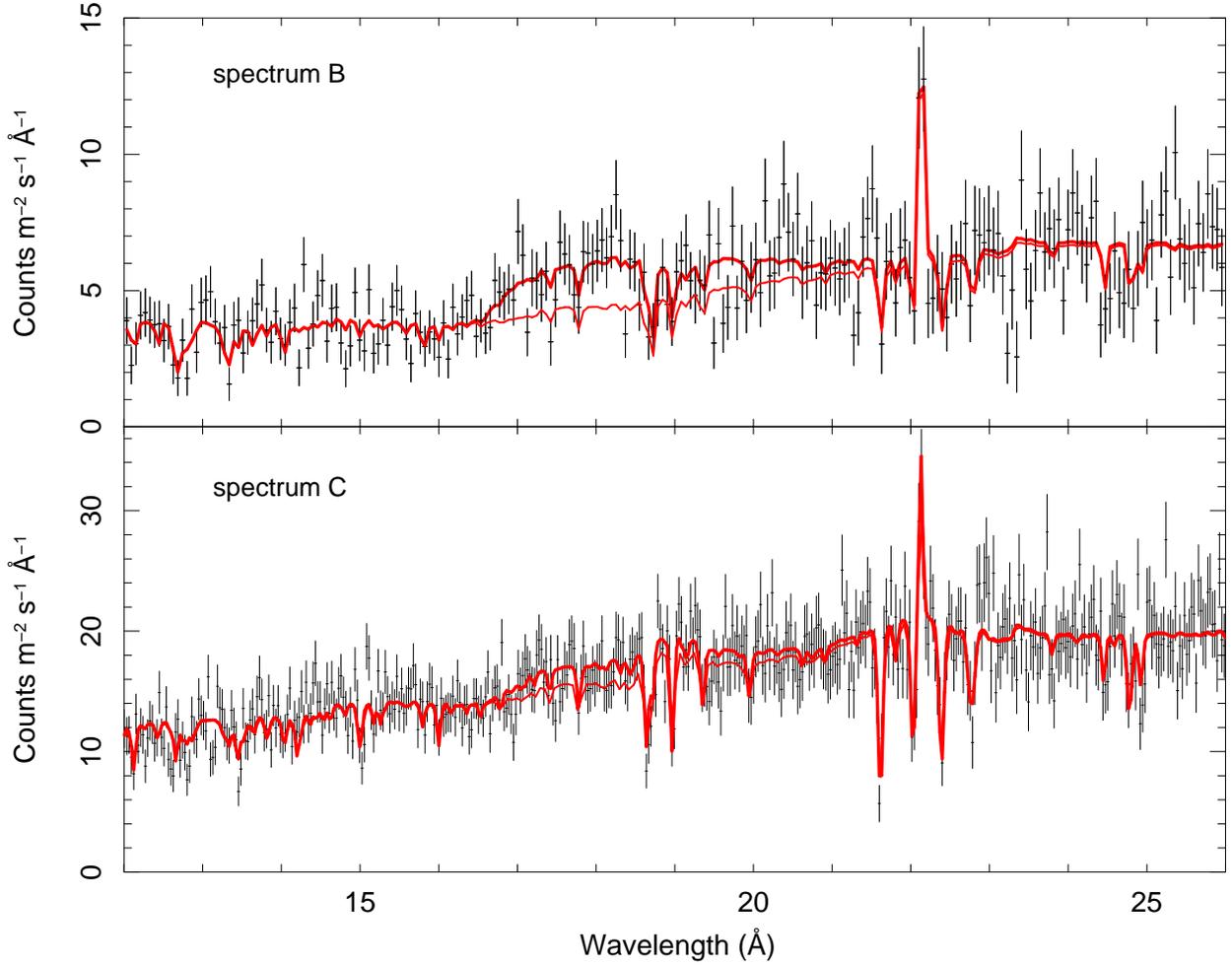}
\caption{Detail of spectra B and C with the best fit model (thick
  line) and the best fit model without the \ion{O}{viii} Ly$\alpha$
  relativistically broadened emission line (thin line).}  
\label{broad_data}
\end{center}
\end{figure*}

\begin{figure}[!htb]
\begin{center}
\includegraphics[angle=-90,width=0.45\textwidth]{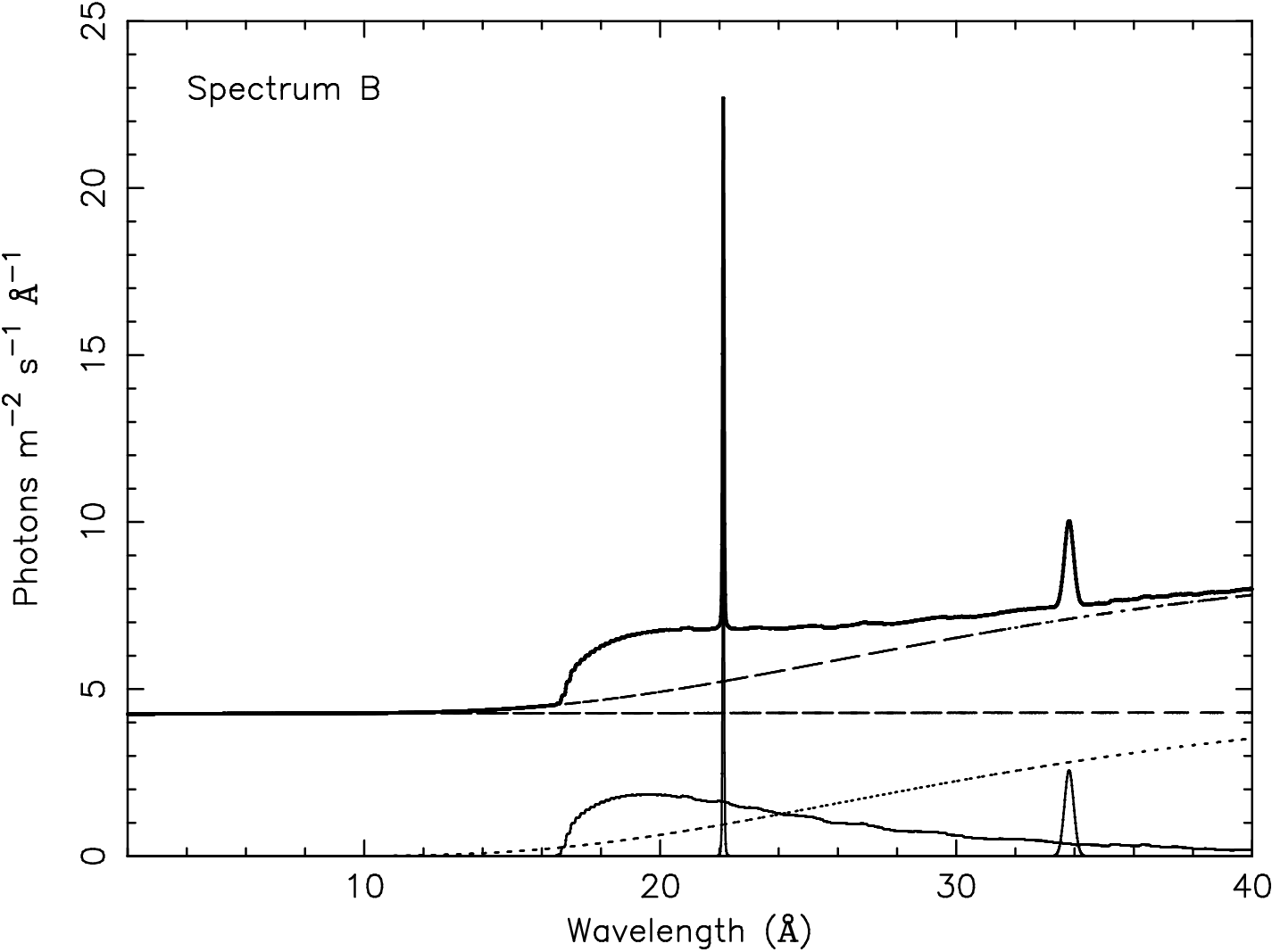}
\includegraphics[angle=-90,width=0.45\textwidth]{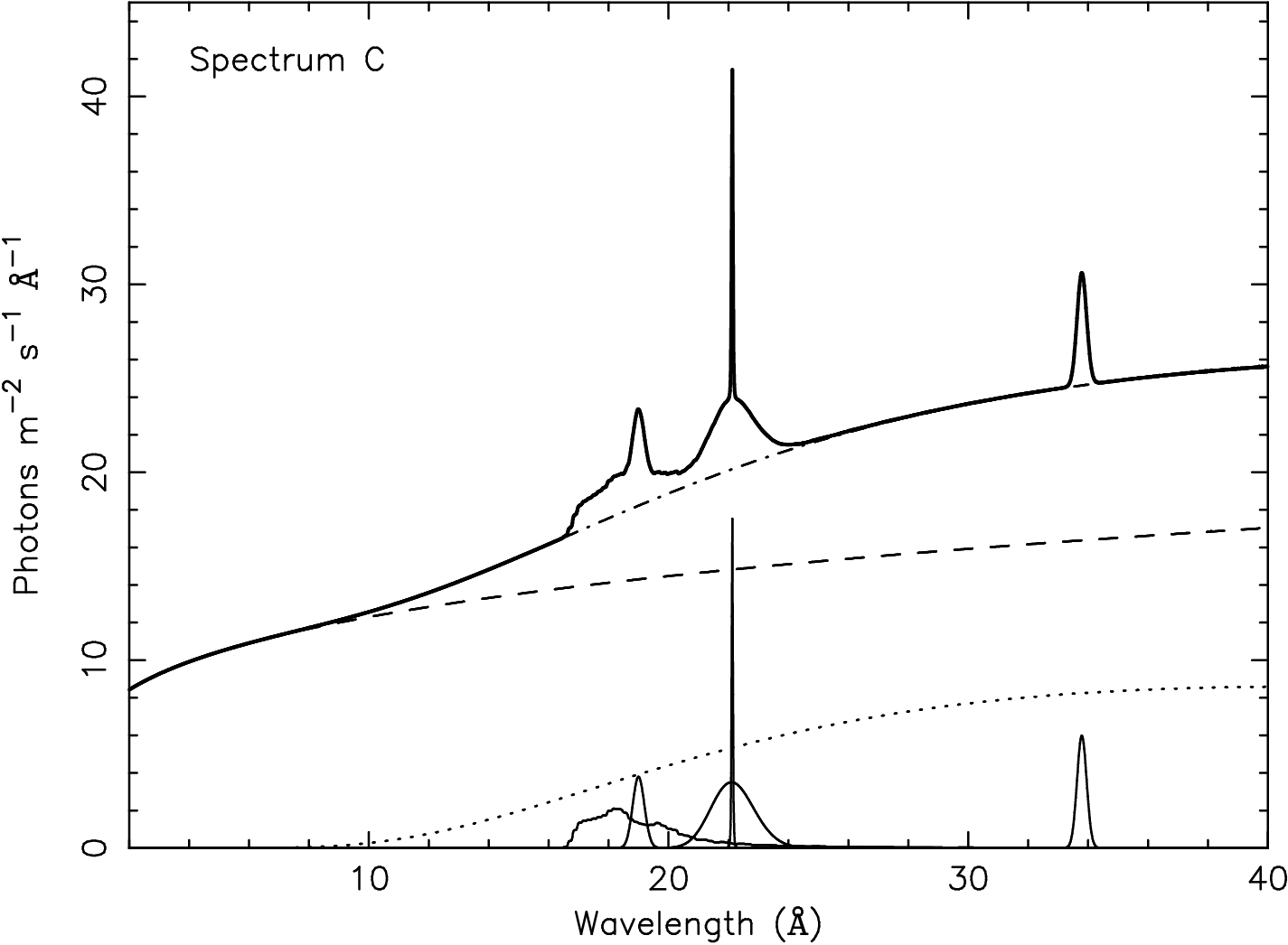}
\caption{Model emission spectra for spectrum B and C, without Galactic
absorption and the warm absorber components. Dashed line: power law
continuum; dotted line: modified black body continuum; dash-dotted
line: sum of power law and modified black body continuum; thin solid
lines: relativistic, broad and narrow emission line contributions;
thick solid line: total unabsorbed emission spectrum.}
\label{broad_line}
\end{center}
\end{figure}

Besides absorption lines, strong emission features are present in the
four spectra. We chose two different methods for fitting this
  excess emission: a model with broadened emission lines and a model
  in which the absorbers only partially cover the continuum. We first
  describe the model with broadened emission lines. As was the case for fitting the absorber, we first
fitted spectrum C and then used its best fit values as a starting
point to fit the other spectra (see Table~\ref{tab:fit} for best fit
parameter values).  The presence of a broad \ion{O}{viii} Ly$\alpha$
emission feature around 19 \AA\ is illustrated in
Fig.~\ref{broad_data}, which shows a detail of the spectra B and C
including the best fit model with and without relativistically broadened
\ion{O}{viii} Ly$\alpha$.  Fig.~\ref{broad_line} shows the model
emission spectrum for spectra B and C. Excess emission with at least a
1 \AA~width, due to the \ion{O}{vii} triplet near 22~\AA~is present in
spectra A, C and D. In addition, there is significant excess emission
in spectra C and D near 34~\AA{}, the region of the 1s--2p transition
in \ion{C}{vi}.

We modelled the excess emission at 19~\AA\ (see Fig.~\ref{broad_line}) with a
relativistically broadened line, following \citet{ogle} and \cite{branduardi}.
The model used is a narrow emission line convolved with the relativistic disc
line profile of \citet{laor}. The inner ($r_1$) and outer ($r_2$) disc radii
were fixed in our models to the default values ($r_1=1.234GM/c^2$ and
$r_2=400GM/c^2$). The rest-frame wavelength of the line was frozen to
18.969~\AA, the wavelength of \ion{O}{viii} Ly$\alpha$.

We have tried fits with different disc inclinations $i$, but in all
fits the inclination converged to 48$^\circ$, within a few
degrees. This value is consistent with the one found by
\citet{ogle}. The disc inclination is not expected to change
significantly over a couple of years. For that reason, and in order to
reduce the number of free parameters, we have kept the inclination
frozen to that value in all presented fits. The disc emissivity slope
$q$ (with $R^{-q}$, $R$ the distance to the black hole) was allowed to
vary, however.

In addition to a relativistic \ion{O}{viii} Ly$\alpha$ line, we added
a Gaussian-shaped broad \ion{O}{viii} line Ly$\alpha$ line (FWHM
0.45~\AA) to our model for spectrum C and D. Further broad lines
corresponding to the \ion{O}{vii} triplet and \ion{C}{vi} Ly$\alpha$
line were also needed. For \ion{C}{vi} Ly$\alpha$, we have frozen the
wavelength to the rest wavelength of 33.736~\AA. For \ion{O}{vii} we
kept the wavelength as a free parameter: the line is a blend of the
\ion{O}{vii} triplet, and the relative contributions of these lines
depend on the physical conditions in the source. Such broadened
emission lines were first detected in NGC~5548 by \cite{kaastra1}, and
confirmed for that source by \cite{steenbrugge}, and were
  detected in this source by \cite{ogle}.

\subsection{Partial covering absorption model}

\begin{figure*}[!htb]
\begin{center}
\includegraphics[angle=-90,width=0.90\textwidth]{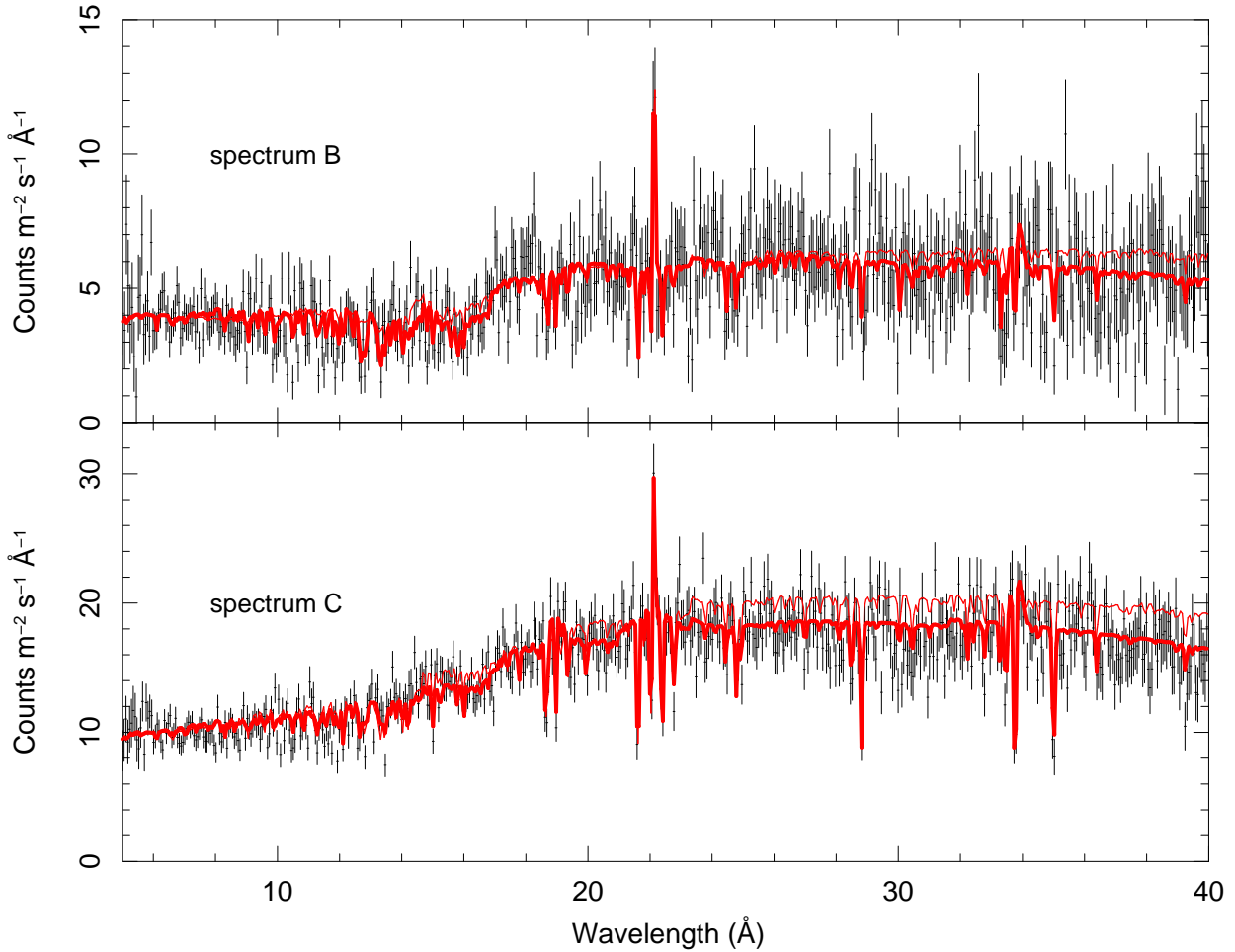}
\caption{{\it Top:} detail of the spectrum for the best fit (thick
  line) model with broad emission lines and the best fit (thin lines) assuming the
  absorber has a covering factor of .75 for spectrum B. {\it Bottom:}
  same as above but for spectrum C.}  
\label{fig:partial}
\end{center}
\end{figure*}

As an alternative to fitting extra emission lines, we fitted the
spectra assuming that the absorber does not fully cover the continuum
emission. Considering that the excess emission is most significantly
detected in spectrum B and C, we first fitted these two spectra (see
Table~\ref{tab:fit} and Fig.~\ref{broad_data}). In this fit there are
no broad emission lines fitted. We fixed the covering factor for the 3
or 4 detected absorption components (spectrum B and C, respectively)
to 0.75, and then refitted the continuum parameters (power law and
modified black body) and the hydrogen column density and ionization
parameter of the detected absorber components. Afterwards we also
allowed the covering factor for each absorption component free to vary
in the fit.

For spectrum C the best fit covering factors for three of the four
absorption components is 1. For warm absorber component 2 the best
fit covering factor is 0.7 $\pm$ 0.13, with an increased hydrogen column density
of 0.63~$\times$~10$^{25}$~m$^{-2}$ (compared with 0.29~$\times$~10$^{25}$~m$^{-2}$ for a covering factor of 1). Forcing all covering factors to 1 increases
$\Delta$C by 17. However, the best fit with partial covering increases
$\Delta$C by 70 for a decrease of 7 in degrees of freedom, compared
with the best fit model with broad emission lines (4 extra free
parameters in the partial absorption fit compared to 11 in the model with broad emission lines). A comparison
between the best fit with broad emission lines and a fit with all
absorbers having a covering factor of 0.75 can be seen in the
bottom panel of Fig.~\ref{fig:partial}. Note that the continuum
long-ward of $\sim$24 \AA~is consistently over-estimated in the fit
assuming that the absorber only covers part of the continuum emission.
 
For spectrum B we followed the same procedure, but here the best fit
for the covering factors of the three absorption components detected
are 1, even if we redid the fitting starting from a covering
factor of only 0.25. The best fit without broad emission lines
increases $\Delta$C by 21 and is shown in
Fig.~\ref{broad_data}. Forcing the three absorbers to have a partial
covering factor of 0.75 worsens the fit by a further 35 in
$\Delta$C. This last fit is compared with the best fit with broad
emission lines in the upper panel of Fig.~\ref{fig:partial}. Note that
again the continuum is over-estimated at long wavelengths, furthermore
the model assuming partial covering cannot reproduce the steep rise at
16 \AA.

\subsection{Narrow emission lines\label{Narrow features}}

The \ion{O}{vii} forbidden line at 22.10 \AA{} is detected with at least
$4\sigma$ significance in each individual spectrum. There is no
  indication for significant flux variability or velocity broadening
  for this line.

In spectrum D a strong, $\sim5\sigma$ significant, RRC from
\ion{C}{vi} (rest wavelength 25.30 \AA) and a $\sim2\sigma$
significant RRC of \ion{C}{v} (rest wavelength 31.62 \AA) are present
(see Fig.~\ref{fig:spectrum_ABCD} and Table~\ref{tab:fitrrc}).  In
addition, we have determined limits to the strength of other expected
RRCs of oxygen and nitrogen ions (see Table~\ref{tab:fitrrc}). The
temperature of the recombining photoionised gas is low but not well
determined ($T$ = $5\pm 2$~eV). In our final fit we fix the
temperature to 5~eV, in order to better assess the significance of
this component. There is a strong correlation between the adopted
temperature and the strength of the emission measure. The RRCs show a
redshift with respect to the rest frame of the galaxy of $\sim
1300$~km\,s$^{-1}$.

\section{Discussion}

\subsection{Continuum}

\begin{figure}[!htb]
\begin{center}
\includegraphics[angle=-90, width=0.45\textwidth]{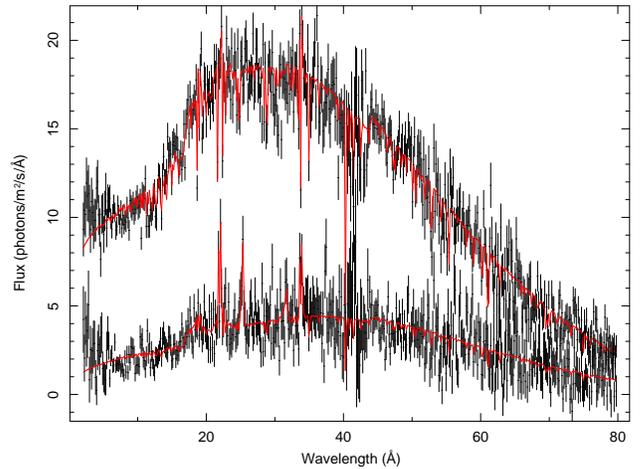}
\caption{The spectrum for part C and D, clearly showing the difference
in continuum shape. The difference is due to a lower flux in the
power-law component and a lower temperature for the modified black
body component in spectrum D. }
\label{fig:cont}
\end{center}
\end{figure}

The uniqueness of the transition between part C and D is that the rather abrupt
flux change is preceded and followed by long periods of relative
quiescence. This allows us to study the response of the continuum, the broad
and narrow emission lines, and the warm absorber to this sudden flux drop. 
We first describe the difference in the continuum. Fig.~\ref{fig:cont}
shows the difference in overall continuum profile for spectrum C and D.

Between spectrum C and D the temperature measured for the modified
black body significantly decreases, and the power-law slope becomes
softer. The temperature remained constant and the power-law slope
softened between spectrum A and B. For Seyfert 1 galaxies, normally,
the spectral slope hardens 
with decreasing luminosity with little or no change in the black body
component.  \cite{pounds} studying two EPIC spectra taken in a high
and a low state of NGC~4051 found that the higher temperature black
body component disappeared in the low state.  The authors found that
the power-law slope did not vary between the high and low state, which
is different from our results.

Changes in the warm absorber properties could mimic a temperature
change of the modified black body component. To ensure that in the spectra
there really is a temperature change we did the following tests. We
refitted spectrum D, but freezing the temperature of the modified
black body to the best fit value of spectrum C.  In the fit the
column density and ionisation parameter of the four absorption
components were left free to vary. The fit thus obtained was
substantially worse (increasing C-statistics by 102), and the fitted ionisation 
parameters were very similar to those obtained in the original fit of
spectrum D. Afterwards, we lowered the ionisation parameter of the only
significantly detected warm absorber component in spectrum D by a
(linear) factor 5. This decrease matches the decrease in luminosity
observed. Leaving the ionisation parameters of the absorber and the temperature
of the modified black body free in the fit,we again derive our
original fit parameters given in Table~\ref{tab:fit}.

\cite{ogle} studying the XMM-{\it Newton} RGS spectrum, find that a
multi-temperature disc model is a poor fit to the soft excess, and
that instead it is well fitted by the relativistically broadened
\ion{O}{viii} line series. We tested this model on our highest
signal-to-noise spectrum, C, by adding to the \ion{O}{viii} Ly$\alpha$
line also the Ly$\beta$, Ly$\gamma$ and the RRC emission lines
relativistically broadened. As none of these emission features are
obvious from a visual inspection, we fixed the relative normalisation
of the different lines to the Ly$\alpha$ component: 0.25 for
Ly$\beta$, 0.1 for Ly$\gamma$ and 0.21 for the RRC. The parameters $q$
and $i$ were free in the fit, but the same for the four different
emission lines. The fit obtained was significantly worse (increasing
C-statistics by 195) for the same number of parameters than a fit with
a modified black body.

We conclude that there is emission from a modified black body present in
the LETGS spectrum, and that the temperature of this component drops
with decreasing luminosity between spectrum C and D.

\subsection{Relativistically broadened \ion{O}{viii} Ly$\alpha$ line} 

The relativistically broadened \ion{O}{viii} Ly$\alpha$ line which
explains part of the RGS soft excess \citep{ogle} is significantly detected in
all spectra (see Table~\ref{tab:fit}). The measured line fluxes do not
follow the power law or modified black body continuum flux
changes. The emissivity $q$, assuming R$^{-q}$, is similar to the
value $q_2 = 5.0$ measured by \cite{ogle} for spectrum B, but not for
spectrum C. We note that for spectrum A and D the emissivity was
frozen to the value for spectrum B and C respectively. \cite{ogle} in detail discusses the steep emissivity derived as well
as the physical parameters in the accretion disc derived from the line
profile and the lack of \ion{C}{vi} Ly$\alpha$ relativistic line.
The latter places a lower limit on the ionisation parameter. Summarized, the
relativistic \ion{O}{viii} Ly$\alpha$ is emitted between 1.6 and 6
gravitational radii, $\xi$ $>$ 3.4 and $n_{\mathrm e}$ $<$ 10$^{23}$ m$^{-3}$. 

The inclination angle for all spectra was consistent with
48$^{\circ}$; very similar to the inclination measured by
\cite{ogle}. The inclination angle for the narrow line region (NLR) as
derived by \cite{christopoulou} from narrow band [\ion{O}{iii}]
imaging and long-slit spectroscopy using the Manchester Echelle
Spectrograph is about 50$^{\circ}$.  This indicates that the narrow
line region, and the gas producing the relativistically broadened
emission line lie along the same line of sight. The narrow line region
is therefore possibly an extension of the gas producing the
relativistically broadened emission lines. If this is the case,
then the gas forming the relativistically broadened emission line is
outflowing, similar to the absorption components. From the kinematics,
\cite{christopoulou} conclude that the imaged [\ion{O}{iii}] gas is
outflowing. The inclination angle of the host
galaxy is at medium angles \citep{mollenhoff}, thus gas observed
from the relativistic and narrow emission lines are closely aligned
with the host galaxy.

Relativistic line profiles in the soft X-ray band were first reported by
\citet{branduardi} in the narrow lines Seyfert 1 galaxies
MCG~$-$6-30-15 and Mrk~766. Interestingly, the relativistically
broadened \ion{O}{viii}, \ion{N}{vii} and \ion{C}{vi} Ly$\alpha$ lines
have only been convincingly detected in narrow line Seyfert 1
galaxies. For MCG~$-$6-30-15 and Mrk~766 the disc line parameters
derived from the soft X-ray band and the Fe-K line are in
good agreement. The XMM-{\it Newton} observation of NGC~4051 that showed a
strong relativistic \ion{O}{viii} line \citep{ogle} did not show a
significant relativistic Fe-K line \citep{pounds}. Unfortunately the
LETGS is not sensitive enough in the the Fe-K band to
directly compare the disc geometry  as derived from iron and oxygen
lines. For the physical implications of the relativistic lines in
NGC~4051 we refer to the discussion by \citet{ogle}. 

\subsection{Broad emission lines}
The broad \ion{O}{vii} line, centred on the forbidden line, is
detected in spectrum C, but the line is weaker or absent in the other
spectra. \cite{ogle} detected a broadened \ion{O}{vii} emission line
with a full width half maximum (FWHM) of 11000 $\pm$ 3000 km
s$^{-1}$. This is consistent with the FWHM we derive: 24000 $\pm$ 6000
km s$^{-1}$. Both values are rather larger than the FWHM determined
from the H$\beta$ and \ion{He}{ii} broad lines measured in the UV
\citep{peterson}.  If the broad line region (BLR) has a stratified ionisation structure, then it is expected that the higher ionised
\ion{O}{vii} line will have a broader width than either the H$\beta$
or lowly ionised \ion{He}{ii}. At 1.7~\AA~the width of the
  broadened \ion{O}{vii} line  exceeds the separation (0.5~\AA)
  between the forbidden and resonance line. The broadened resonance (and
intercombination) lines contribute to the flux and hence the width
measured.

We also detect a broadened \ion{C}{vi} line with a width of 3200 $\pm$ 1000 km
s$^{-1}$, which  is smaller by a
factor of $\sim$7 than the \ion{O}{vii} line. This value is fairly similar to the FWHM derived for the
H$\beta$ line (1100 $\pm$ 190, \citealt{peterson}) and the width
measured for this line in the RGS spectrum (1200 km s$^{-1}$,
\citealt{ogle}). The smaller width is consistent in terms of a stratified
BLR model, with the fact that \ion{C}{vi} is formed at lower
ionisation parameters.

In spectrum C on top of the relativistically
broadened \ion{O}{viii} Ly$\alpha$ emission line there is a broadened
\ion{O}{viii} Ly$\alpha$ emission line.

\cite{costantini} studying Mrk~279 did detect broadened emission lines
with better statistics than is the case for NGC 4051.  The authors
conclude that the broad X-ray emission lines originate from a
stratified BLR. The above findings for NGC~4051
are consistent with this conclusion.

\subsection{Narrow emission features}

During our observations we detect various narrow emission features in the
spectra. The narrow forbidden \ion{O}{vii} line was significantly
detected in all spectra, and all flux values are 
consistent with each other to within 1$\sigma$ (see Table~\ref{tab:fit}). This is
consistent with the picture that the \ion{O}{vii} forbidden line is formed at the narrow
line region further out. The width of this line (it is unresolved), precludes it being formed as part of the outflow. 

In spectrum D we detect the \ion{C}{vi} RRC and the \ion{C}{v}
RRC. In the other three spectra these features are not
detected. The shapes and centroids of individual RRCs are determined by
the temperature of the recombining gas, the emission measure of the
relevant ion, and by the motion (velocity broadening) of the
recombining gas. We derived the temperature and emission measure for
the RRCs assuming no velocity broadening. We have investigated the
effect of broadening by including a velocity
broadening component in our RRC spectral model. On the basis of the goodness of fit we exclude
that the turbulent velocity in the recombining gas is greater than
$3\,000\,\mathrm{km\,s}^{-1}$. Both the temperature and emission measure
are within the error bars the same for models with and without a
velocity broadening of $3\,000\,\mathrm{km\,s}^{-1}$.

If we interpret the maximum velocity broadening as due to Keplerian
velocity at the location where the RRC is formed and assume a black hole mass
of $M=3\times10^5 M_\odot$ \citep{mchardy} we obtain a minimum
distance for the recombining gas of $R\sim 4\times 10^{12}$ m, or $\sim 4500$ Schwarzschild radii.

From the fitted temperature of the RRC of $T$ = $5 \pm 2$~eV, we derive an
ionisation parameter $\log \xi \simeq 1.6$. With the ionising
luminosity of the source ($L=2.8\times 10^{35}$ W), the minimum
distance derived above and the definition of $\xi$, we obtain a maximum
density of 4$\times$10$^{17}$~m$^{-3}$. It is possible to estimate a
lower limit to the volume of the RRC emitting source using the
observed ionic emission measure of the \ion{C}{vi} RRC ($3.9\times
10^{66}$~m$^{-3}$), a solar carbon abundance \citep{anders89} and a
fraction of 60~\% in the form of \ion{C}{vii} (appropriate for
$\log\xi = 1.6$).  We find an emitting volume of
$V=10^{35}$~m$^{3}$. Note that the volume might be larger than
the emitting volume if the filling factor is smaller than unity. We
should note that CLOUDY \citep{ferland} version 95.06 calculates the
size of the clouds to be of order 10$^{11}$ m, thus a filling factor
of less than unity is rather unlikely. We equate this volume to $\Omega
R^2\Delta R$ where $\Delta R$ the characteristic thickness of the
emitter in the radial direction and $\Omega$ the solid angle sustained
by the emitter.  Dividing the above formula by $R^3$ and substituting
the values for the volume and minimum distance we find:
\begin{equation} 
\Omega \frac{\Delta R}{R} = 0.002.
\end{equation}

Taking for the distance a value of 4 $\times$ 10$^{13}$ m,
  instead of 4 $\times$ 10$^{12}$ m, we derive a value of 0.02.

Another constraint follows from the column density $N_{\mathrm H}$
through the emitting region. This can be written as $N_{\mathrm H} =
n\Delta R \leq n R (0.002 / \Omega)$ (assuming the filling factor
  is unity, see Eq. 1). Taking the
minimum radius and maximum density we can get an indication of the
column density: $N_{\mathrm H} \sim
3\times10^{26}(4\pi/\Omega)$~m$^{-2}$.

A possible origin of the RRCs is that the gas in one (or more) of the
warm absorber components is recombining. To test this hypothesis, we
compare the properties of the recombining gas with that of the warm
absorber components. The large estimated column density for the RRC is
only derived for absorption component 4. However for this component
the ionisation parameter is substantially higher, making a connection
between this absorption component and the gas producing the RRCs
unlikely. Component 2 has a similar ionisation parameter to the RRCs
but its column density in spectrum B and C is substantially
smaller. Thus this connection is also unlikely. Alternatively the
RRCs could originate in the thin, ionised skin of the accretion
disc. The width of the \ion{C}{vi} broad emission line is similar
  to the maximum broadening of the RRCs. Also the ionization
  parameters are similar. 

For $\log \xi \simeq 1.6$ we expect that most of the oxygen gas will recombine
into \ion{O}{vii}, thus we expect a strong \ion{O}{vii} RRC at 16.77
\AA. This RRC is not observed, and we can place an upper limit 
for the emission measure of 0.6 $\times$ 10$^{66}$
m$^{-3}$. It is possible that, the many absorption lines due to different iron
ions mask the presence of the \ion{O}{vii} RRC, or that the temperature is
lower than 5 eV.

\citet{ogle} detected the \ion{C}{vi} RRC in the high flux state RGS
spectrum; \citet{pounds} detected in a different low state RGS spectrum
the \ion{C}{vi} and \ion{O}{vii} RRCs. Both authors derive a
temperature of 3 eV, consistent with our derived temperature of 5
$\pm$ 2 eV.

\subsection{The ionised absorber}

In our spectra, we have found evidence for the presence of four warm
absorber components (Table~\ref{tab:fit}). The absorption components
are discussed in detail below. 

\subsubsection{Warm absorber component 1}

The least ionised component, which has a low outflow velocity and the
smallest hydrogen column density, is not detected in spectrum A or B,
due to the lower signal to noise in these spectra. The strongest lines are
due to \ion{O}{iv}, \ion{O}{v} and \ion{C}{v}.

\subsubsection{Warm absorber component 2}

This component has a similarly low outflow velocity as component 1,
but with a higher ionisation parameter, log $\xi$=0.87. The strongest
lines of this component are the deep 1s--2p absorption lines of
\ion{O}{vii}, \ion{O}{vi} and \ion{O}{v}. This is the only component
for which we measure a decrease by a factor of two in the ionisation parameter
between spectrum C and D (see Table~\ref{tab:fit},
Fig~\ref{fig:stepad}). This component, perhaps partly together with
our component 1, corresponds to the low ionisation component found by
\citet{krongold}. However, they find a column
density that is twice as large and a somewhat higher outflow velocity of
$-500$~km\,s$^{-1}$. 

\subsubsection{Warm absorber component 3}

With an outflow velocity of $-$580 km s$^{-1}$, component 3 has
the largest opacity. The dominant imprints are the Ly$\alpha$
transitions of hydrogenic oxygen, nitrogen and carbon. The outflow
velocity and ionisation state are similar to one of the two absorber
components reported by \cite{collinge}. Note that the observed Doppler
velocity (sum of the $+700$~km\,s$^{-1}$ heliocentric velocity of
NGC~4051, \citet{verheijen}, and the intrinsic outflow velocity of
$\sim -600$~km\,s$^{-1}$) is only $+100$ km~s$^{-1}$, and less than
2$\sigma$ different from 0. Therefore, this absorption could be blended with
absorption from our Galaxy or the local group. \cite{pounds} analysed
the high state RGS spectrum; and  detected an absorber with log $\xi$=2.7 and
an outflow velocity of $\sim -$600 km~s$^{-1}$, which can probably be
associated with this component. We associate this component with
the high ionisation component of \citet{krongold}, for which
these authors derive a distance of $1-2\times 10^{13}$~m based on time
variability; but see Sect.~5.5.6 below.

\subsubsection{Warm absorber component 4}

The last absorber component has the highest ionisation parameter,
hydrogen column density and outflow velocity. The \ion{C}{vi} and
  \ion{N}{vii} Ly$\alpha$ lines are the deepest detected. The \ion{O}{viii}
  Ly$\alpha$ blends with and broadens the less blueshifted secondary
  \ion{O}{vii} line. The outflow velocity of
$-$4670 km s$^{-1}$ was first reported by \cite{vandermeer}. This is
one of the highest outflow velocities derived from high resolution
spectra of Seyfert galaxies, that we are aware
of. \cite{collinge} observed for the highest ionised component in a
{\it Chandra} HETG spectrum of NGC~4051, an outflow velocity component of $-$2340
km s$^{-1}$, 
which is not detected in our spectra. The high velocity component that
we
detect, and the one reported by \cite{collinge}, seem to be the only
ionised components detected in the X-rays without a UV absorber at the
same outflow velocity. In many AGN the most highly ionised X-ray absorber
should not produce any UV absorption lines, however usually a UV
absorber with the same outflow velocity as the X-ray absorber is detected. This does not seem to be the case for the
$-$4670 km s$^{-1}$ component we detect.  Both the $-$2340 km s$^{-1}$ and the
$-$4670 km s$^{-1}$ components are highly ionised and therefore detectable with
the HETG, the LETG and the RGS. \cite{ogle} do not detect these highly
ionised components in the RGS spectrum. Considering that neither we nor
\cite{ogle} detect the $-$2340 km s$^{-1}$ component, and that the $-$4670
km s$^{-1}$ component is not present in the earlier {\it Chandra} HETG
and RGS data, these high velocity absorbers either have a variable column
density or are transient.

\subsubsection{Comparison between the absorber components}
 
The derived hydrogen column density and ionisation parameters for our
three absorbers with an ionisation parameter log $\xi < $3 are in
agreement with those derived from an RGS observation, in a
similar flux state, and given in a figure by \cite{ogle}. We confirm,
using the parameters measured for spectrum C, that there is an increase in
hydrogen column density with increasing ionisation parameter, as
observed in this source by \cite{ogle} and in NGC~5548 by
\citet{steenbrugge}.

It is also possible to constrain the opening angle of the outflow. From the ionising
luminosity ($L=2.8\times 10^{35}$~W) and the definition of $\xi$, we get
$nr^2=L/\xi$. As we measure the outflow velocity $v$, we know that the mass
outflow rate per steradian is $m_{\mathrm p}nr^2v$. We list this value for each
of the absorber components in Table~4, based on our result for
spectrum C. Assuming a typical accretion efficiency of 10~\%, the
observed luminosity of NGC~4051 corresponds to an accretion rate of
$5\times 10^{-4}$~M$_{\odot}$\,yr$^{-1}$. Assuming the mass
outflow through the wind is smaller than the accretion onto the black
hole, we can put upper limits to the solid angle sustained by
the outflow. This assumption seems reasonable as the energy
  needed to accelerate the outflow must come from the infalling gas.
These upper limits are small,
typically less than $10^{-2}$ to $10^{-5}$~sr. Using this same
formalism, similar to somewhat larger opening angles were derived
for the outflow in NGC 5548 \citep{steenbrugge}.

\begin{table}[!htbp]
\begin{center}
\caption{Observed and calculated parameters for the four warm absorber components of spectrum C.}
 \begin{tabular}{@{\extracolsep{-2.5mm}}lcccc}
\hline
\hline
component & 1 & 2 & 3 & 4 \\
\hline
$\log\xi$ & 0.07 & 0.87 & 2.32 & 3.19 \\
$\log T$ (K) & 4.21 & 4.33 & 5.14 & 5.83 \\
$nr^2$ (m$^{-1}$) & $2.4\times 10^{44}$ & $3.8\times 10^{43}$ 
                  & $1.3\times 10^{42}$ & $1.8\times 10^{41}$ \\
$v$ (km\,s$^{-1}$) & -210 & -200 & -580 & -4670 \\
$m_{\mathrm p}nr^2v$ (M$_{\odot}$\,yr$^{-1}$) & 1.3 & 0.20 & 0.021 & 0.022 \\
$\Omega$ (sr) & $<$3.8$\times$10$^{-4}$ & $<$2.5$\times$10$^{-3}$ &
$<$2.4$\times$10$^{-2}$ & $<$2.2$\times$10$^{-2}$ \\\hline 
most abundant & & & & \\
oxygen ion $i$ & \ion{O}{v} & \ion{O}{vii} & \ion{O}{viii} & \ion{O}{viii} \\
$\tau_i n$ (s\,m$^{-3}$) & $5.3\times 10^{16}$ & $8.8\times 10^{16}$ 
                         & $8.5\times 10^{16}$ & $7.4\times 10^{15}$ \\
\multicolumn{5}{l}{For $\tau_i<20$~ks or $\tau_i>20$~ks:}\\
limit on $r$ (m) & $>9\times 10^{15}$ & $<3\times 10^{15}$ 
                 & $>5\times 10^{14}$ & $>7\times 10^{14}$ \\
\multicolumn{5}{l}{For $\tau_i<570$ days:}\\
limit on $r$ (m) &        --          & $<2\times 10^{17}$ 
                 & $<3\times 10^{16}$ & $<4\times 10^{16}$ \\
\hline
\end{tabular}
\label{tab:par}
\end{center}
\end{table}

\subsubsection{Time variability of the warm absorber}

Our spectral modelling of the warm absorber components is based upon models for
photoionisation balance. Whenever the ionising flux of the source changes, the
ionisation balance should adjust to the new situation, and the ionisation
parameter $\xi$ should change. The speed at which this happens depends on the density $n$
of the absorbing material. The recombination timescale $\tau_i$ for a given ion
$i$ is inverse proportional to the gas density. It is given by
\cite{bottorff} to be:

\begin{equation}
\tau_i n = \bigl[ 
\alpha_i ( \frac{f_{i+1}}{f_i} - \frac{\alpha_{i-1}}{\alpha_i} )
\bigr]^{-1}
\label{eqn:taurec}
\end{equation}
where $f_{i}$ is the relative concentration of ion $i$ of a given element,
and $\alpha_i$ is the recombination rate of ion $i$, and in general depends
only on the temperature $T$ of the gas. This formula does take
  into account the cascade into and out of ion $i$. Furthermore, a
  negative value for the recombination time indicates that the ion is
  destroyed by recombination to the ion $i$-1. For a given photoionisation equilibrium,
the temperature is a unique function of the ionisation parameter $\xi$,
and the ion concentrations are also known as a function of $\xi$. Therefore,
given a value for $\xi$, the right hand side of (Eq.~\ref{eqn:taurec}) can be
evaluated.

The largest flux change occurs between spectrum C and D. The hard $2-10$~keV power-law flux drops suddenly by a factor of 6.1 ($\log p=-0.79$,
where we define $p$ as the flux ratio of spectrum D to spectrum C). The
soft flux (taken here and later as the $0.2-10$~keV modified black body flux
from Table~1) decreases by a factor of 5.4 ($\log p=-0.73$). Thus, if the
density in one of the components is high, we would expect $\log\xi$ to decrease
by about $0.73$, and if the density is low, we expect no change between
spectrum C and D. As the low state of spectrum D lasts about 20~ks and
our data have insufficient S/N to resolve spectrum D further in time,
this gives the limit on the recombination timescale. 

Indeed in components 1, 3 and 4 we see no significant change in
$\log\xi$, and definitely no change as large as $-0.73$ (see Table~1),
indicating that the recombination timescale is longer than 20~ks. This
then gives the lower limits to the distance of these components from
the central source as listed in Table~\ref{tab:par}. In teh table the
product $nr^2$ is calculated from $L/\xi$. The product of
recombination timescale and density $\tau_i n$ is calculated from
(Eq.~\ref{eqn:taurec}), using the total recombination rates of
\citealt{nahar}.  Limits on $r$ are calculated from
$r=\sqrt{(nr^2)(\tau_i)/(\tau_i n)}$. The lower limit for component 3,
namely $5\times 10^{14}$~m (19 light days), disagrees with the
distance of $0.5-1$ light days derived by \citet{krongold} for this
component.

\cite{krongold} studied the variability using the
relatively low-resolution EPIC data. Some possible reasons for the
disagreement are given below. \cite{krongold} do not allow the
continuum shape (i.e. photon index and temperature) to vary, while at
least in our data the shape of the continuum significantly changes
(see Fig.~\ref{fig:cont}). Further, \cite{krongold} do not include a
relativistic \ion{O}{viii} Ly$\alpha$ line although in the total and
high state spectra their best fit model (see their figures 2 and 3) is
consistently below the observed continuum between 17 and 19.2
\AA. Similarly, the continuum is above their model between 20 and
21.8 \AA, where we fit a broadened \ion{O}{vii} line. Failure to fit the
local continuum correctly will result in too low or too high column
densities derived from the line depths. 

\cite{krongold} included only
two absorption components and have problem fitting some of the oxygen
lines. Therefore, their model cannot explain the \ion{O}{v} absorption
line at 22.4 \AA, which is easily observed in their published RGS
spectrum.  Furthermore, the depth of the \ion{O}{vii} line is
overestimated in the high state, and underestimated it in the low
state. Oxygen lines are diagnostically important because of their
large oscillator strength (i.e. strong lines) and the rather large
range in ionisation parameters they span. One can thus determine (over
this ionisation range) the ionisation structure without having to worry
about abundance effects. Finally, long term differences between our
observations and those of \cite{krongold}, could explain the
differences found in ionisation variability behaviour. 

\cite{krongold} claim to detect variability in the ionisation
parameters on timescales of 2-3~ks, leading to a much larger derived
density and thus closer location of the absorber.

In our component 2, we see a decrease of a factor of 2
($\Delta\log\xi = -0.35\pm 0.16$), suggesting that the recombination
time is indeed 20~ks or smaller. This now gives an upper limit to the
distance of $3\times10^{15}$~m or 110 light days (see Table~\ref{tab:par}). This
is higher than, but consistent with the upper limit of 3.5 light days
for the low ionisation component of \citet{krongold}. That the change
in $\log\xi$ is not the expected value of $-0.73$ may have two reasons,
which we cannot be distinguished in our present data. The recombination
timescale might be close to our upper limit, so that the gas is still
recombining at the end of interval D. Alternatively, the flux changes
in the invisible EUV band might be smaller, and these may be more
important for the low ionisation gas than the X-ray continuum (the dominant
oxygen ions for component 2 are \ion{O}{vi} and \ion{O}{v}).

Next we investigate the long term variability, by comparing spectrum B to
spectrum C in a similar way as we did above for spectra D and C. The soft and
hard flux of spectrum B have $\log p = -0.48$ and $\log p = -0.33$,
respectively. Taking a typical value of $-0.4$, we see that all visible
components in spectrum B (components $2-4$) are in agreement with
such a change in $\log\xi$ ($\Delta\log\xi = -0.38 \pm 0.22$, $-0.34
\pm 0.26$ and $-0.27 \pm 0.19$). We conclude that the
recombination timescale for 
these components is likely faster than 570 days, the interval between both
observations. This gives interesting upper limits to the distance of components
3 and 4 (see Table~4) of about 1~pc. We should note that
  components 2, 3 and 4 could coexist in the same location, but that
  component 1 and 2 can not coexist in the same location, due to the
  lack of variability in component 1, even over 570 days.

Finally, we have an independent short term comparison by comparing
spectrum A to spectrum B, although our discriminating power is smaller
here, because of the smaller flux jump ($\log p = -0.13$ and $-0.21$
for the soft and hard band, respectively). The error bars on the
ionisation parameter for components 3 and 4 preclude a definite
conclusion here; for component 2, we see an increase in spectrum A of
$\Delta\log\xi = +0.46\pm 0.32$, which favours the no change scenario,
but does not exclude a small response compatible with $\log
p=-0.13$. We also note that, of all our four spectra, the flux during
spectrum A shows the largest variations within the given interval
(Fig.~1), so the establishment of any (quasi)-equilibrium during that
interval may be questioned.

\section{Summary}

In this study we have investigated the spectral properties of NGC~4051, 
observed by {\it Chandra}-LETGS on two occasions for a total of 180\,ks.
Due to the flux variations within each observation, we 
split each observation into 2 spectra, to be fit separately. In the
highest signal-to-noise spectrum, C, we detect four absorption
components. Our main results are:

\begin{itemize}
\item The continuum parameters do change between the different observations,
the power-law slope becomes softer with time. The temperature of the
modified black body increases, to first order, with increasing
luminosity. 

\item Between spectra C and D, when the flux dropped by a factor of 5,
  absorbers 1, 3 and 4 show no evidence 
for a change in ionisation parameter; however, for absorber 2 the
ionisation parameter does decrease.

\item Our limits on the variability of the absorber suggest that component
2 is located within 0.1~pc, and components 3 and 4 are located somewhere in the
range $0.02-1$~pc. For component 1, we only have a lower limit of
0.3~pc. These distances are different from those derived by
  \cite{krongold}.

\item We detect one of the highest outflow velocities ($-$ 4670 km
s$^{-1}$) for an ionised absorber observed in the soft X-ray band. We
do not confirm the $-2340$\,km s$^{-1}$ absorber detected by
\cite{collinge}, indicating that these high outflow velocity absorbers
have variable column densities or are transient.

\item The RRCs of \ion{C}{vi} and \ion{C}{v} are detected in spectrum D,
after a drop in flux by factor 5. These are unlikely due to recombining
gas from one of the absorber components, but are possibly from the
accretion disc. 

\item Broad emission lines of the \ion{O}{vii} triplet and
\ion{O}{viii}~Ly$\alpha$ and \ion{C}{vi}~Ly$\alpha$ are
detected. These lines show a simple Gaussian profile, suggesting an
origin in the BLR.

\item We detect a relativistically broadened \ion{O}{viii}
Ly$\alpha$ line, similar to the findings by \cite{ogle}. 

\end{itemize}

\begin{acknowledgements}
KCS thanks St John's College for a fellowship. SRON
is supported financially by NWO, the Netherlands Organization for Scientific
Research.
\end{acknowledgements}

\end{document}